\documentclass{aa}  
\usepackage{bm}
\usepackage{natbib}
\usepackage{subcaption}
\usepackage{multicol}

\newcommand{\vect}[1]{\mathbf{#1}}
\renewcommand{\vec}{\vect}
\newcommand{\td}[2]{\frac{d #1}{d #2}}

\newcommand{\pd}[2]{\frac{\partial#1}{\partial#2}}

\usepackage{graphicx}
\usepackage{txfonts}
\begin{document}

      \title{Modelling three-dimensional transport of solar energetic protons in a corotating interaction region generated with EUHFORIA}
      \titlerunning{Modelling 3D transport of SEPs with EUHFORIA}   
   \author{N. Wijsen
          \inst{1,2}
          \and
          A. Aran\inst{2}
          \and
          J. Pomoell\inst{3}
          \and
          S. Poedts\inst{1}
          }

   \institute{Department of Mathematics, KU Leuven, Belgium\\
              \email{nicolas.wijsen@kuleuven.be}
          \and
                 Departament F\'{i}sica Qu\`antica i Astrof\'{i}sica, Institut de Ci\`encies del Cosmos (ICCUB), Universitat de Barcelona, Spain
         \and
                 Department of Physics, University of Helsinki,  Finland
             }

   \date{Received 26 July 2018 / Accepted  18 November 2018}

 
  \abstract
   {}
   { We introduce a new solar energetic particle (SEP) transport code that aims at studying the effects of different background solar wind configurations on SEP events. In this work, we focus on the influence of varying solar wind velocities on the adiabatic energy changes of SEPs and study how a non-Parker background solar wind can trap particles temporarily at small heliocentric radial distances ($\lesssim 1.5$ AU) thereby influencing the cross-field diffusion of SEPs in the interplanetary space.}
   {Our particle transport code computes particle distributions in the heliosphere by solving the focused transport equation (FTE) in a stochastic manner.  
   Particles are propagated in a solar wind generated by the newly developed data-driven heliospheric model, EUHFORIA. 
    In this work, we solve the FTE, including all solar wind effects, cross-field diffusion, and magnetic-field gradient and curvature drifts. As initial conditions, we assume a delta injection of 4 MeV protons, spread uniformly over a selected region at the inner boundary of the model. To verify the model, we first propagate particles in nominal undisturbed fast and slow solar winds. Thereafter, we simulate and analyse the propagation of particles in a solar wind containing a corotating interaction region (CIR). We study the particle intensities and anisotropies measured by a fleet of virtual observers located at different positions in the heliosphere, as well as the global distribution of particles in interplanetary space.
    }
   {
The differential intensity-time profiles obtained in the simulations using the nominal Parker solar wind solutions illustrate the considerable adiabatic deceleration undergone by SEPs, especially when propagating in a fast solar wind. 
In the case of the solar wind containing a CIR, we observe that particles adiabatically accelerate when propagating in the compression waves  bounding the CIR at small radial distances. 
In addition, for $ r  \gtrsim 1.5$ AU,  there are  particles accelerated by the reverse shock as indicated by, for example, the anisotropies and pitch-angle distributions of the particles. Moreover,  a decrease in high-energy particles at the stream interface (SI) inside the CIR is observed.  The compression/shock waves and the magnetic configuration near the SI may also act as a magnetic mirror, producing long-lasting high intensities at small radial distances.  We also illustrate  how the efficiency of the cross-field diffusion in spreading particles in the heliosphere is enhanced due to compressed magnetic fields. Finally,  the inclusion of cross-field diffusion enables some particles to cross both the forward compression wave at small  radial distances and the forward shock at larger radial distances. This results in the formation of an accelerated particle population centred on the forward shock, despite the lack of magnetic connection between the particle injection region and this shock wave. Particles injected in the fast solar wind stream cannot reach the forward shock since the SI acts as a diffusion barrier.
   }
   {}


   \maketitle
%

\section{Introduction}

Occasionally, particle monitors on board spacecraft will register sudden strong increases in particle fluxes over a wide range of energies. 
The origin of these particles typically lies in eruptive events occurring at the Sun, and they are therefore commonly known as solar energetic particles (SEPs).
The majority of  these SEPs are expected to have gained high energies either by means of stochastic acceleration mechanisms in a parent flare region or by means of first-order Fermi acceleration at shock waves driven by coronal mass ejections (CMEs). 
Apart from solar eruptive events, corotating interaction regions (CIRs) are considered an important source of energetic particles in the heliosphere \citep{richardson04}. At larger heliocentric
distances, these CIRs are typically bounded by a forward shock wave propagating in the slow solar wind, and a reverse shock wave propagating in the fast solar wind \citep[see e.g.][]{richardson18}. Like the shock wave in front of a CME, the shock waves associated with a CIR can potentially also accelerate particles to high energies \citep{fisk80, classen98}. 

Once particles escape from their acceleration site, they travel through the heliosphere, spiralling around the interplanetary magnetic field (IMF) lines.  
During their journey, particles interact with small-scale magnetic turbulence omnipresent in the solar wind. 
This turbulence can scatter the particles,  and hence determines their mean free path.  
A large number of
studies have been conducted that focus on
quantifying the amount of magnetic turbulence in the solar wind, and studying the effect of different mean free paths on particle events \citep[e.g.][and references therein]{beeck89, kunow91,bieber94}. 
 This has typically been done by assuming a Parker solar wind configuration \citep{parker58} and describing the effect of turbulence on particle transport through diffusive processes in the particle's spatial coordinate or  pitch-angle. 
Aside from the characteristics of small-scale magnetic-field turbulence, energetic particle transport, and hence SEP events, can also strongly  be affected by the global solar wind configuration. 
In reality, the latter is seldom described well by a steady-state Parker configuration, since the characteristics of solar wind source regions, like coronal streamers and coronal holes, are very different from one another, leading to varying solar wind speeds and  densities. 
Apart from that,  transient solar eruptive events strongly affect the conditions in interplanetary (IP) space. Therefore, it is important to include these varying solar wind conditions in SEP transport models, in order to achieve a better understanding of SEP events.
 
In this work we introduce a new energetic particle transport code that aims at studying the effects of  a  solar wind that is more complex than a nominal Parker wind configuration on SEP events.
The new  three-dimensional (3D) particle transport code solves the focused transport equation (FTE) and computes particle distributions in the heliosphere by means of a Monte Carlo simulation.
This code is coupled to a newly developed data-driven heliospheric model that solves the 3D magnetohydrodynamic (MHD) equations in IP space, the EUropean Heliospheric FORecasting Information Asset (EUHFORIA) model \citep{pomoell18}. 
This model allows us to obtain realistic solar wind configurations in which we then propagate the energetic particles.  

Other previous efforts have coupled 3D MHD simulations of the solar wind and CME-driven shocks with particle FTE transport codes with different levels of simplification. 
For example, \cite{gasen11,gasen14} used the Shock-and-Particle modelling approach \citep[e.g.][]{pomoell15,aran08,lario98} to model the propagation of a CME-driven shock from near the Sun (i.e. $4 R_{\odot}$) to 1 AU and coupled it with a transport model  describing the propagation of protons, under nominal upstream solar wind conditions, to describe the variation of SEP event peak intensities with the radial, longitudinal and latitudinal position of the observers. 
The SEPMOD model developed by \cite{luhmann07,luhmann10,luhmann17} describes the scatter-free transport of SEP events generated by CME-driven shocks (from 0.1  AU), in non-uniform solar wind conditions simulated by using the ENLIL model \citep{odstrcil04,odstrcil05}.
The outputs of SEPMOD are available through the Community Coordinated Modelling Center \citep{luhmann17}. 
\cite{kozarev10} also used the ENLIL model to simulate the propagation of particles from the observed intensities at 1 AU to further distances from the Sun during the events in October-November 2003. 
The propagation model they used, the Energetic Particle Radiation Environment Module (EPREM), can be applied to any interplanetary magnetic field (IMF) configuration by solving the perpendicular diffusion and drift separately from the rest of the transport effects included in the FTE equation \citep[][and references therein]{schwadron10}. 
More recently, \cite{kozarev13} coupled the EPREM model with the Block Adaptive Tree Solar-Wind Roe Upwind Scheme (BATSRUS) model \citep[e.g.][]{toth12,manchester12} to study the acceleration and transport of protons in the solar corona during the 2005 May 13 SEP event. 
\cite{schwadron14} introduce the The Coronal-Solar Wind Energetic Particle Acceleration (C-SWEPA) models by simulating the space weather effects of a synthetic extreme SEP event. The SEP transport and acceleration is modelled with EPREM. 
These authors highlight the importance of the particle cross-field diffusion in the longitudinal spreading of particle distributions.

In this work, we employ the standard-drift guiding centre FTE equation \citep[e.g.][]{leRoux09} with the addition of a perpendicular spatial diffusion term. We first briefly illustrate the transport effects for protons travelling in a nominal slow or fast Parker solar wind configuration. 
Subsequently,  we present a detailed study of SEP transport in a EUHFORIA-generated  slow  solar wind with an embedded fast solar wind stream, forming a shock bounded CIR at large heliospheric radial distances ($r\gtrsim 1.5$ AU).
 \cite{Giacalone02} introduced an analytical model for a CIR at small radial distances ($r\sim 1$ AU) where the forward
and reverse shocks have not yet formed, to study its effect on interstellar pickup-ions. 
This analytical model was later used by for example \cite{03kocharov,kocharov08a} to illustrate that corotating compression regions can modify the time-intensity profiles, anisotropies, and energy spectra  of SEP events. 
Both \cite{Giacalone02} and \cite{03kocharov} also illustrated that a forward or reverse compression wave can act as a magnetic mirror, temporarily trapping energetic particles at small radial distances. 
The analytical CIR model of \cite{Giacalone02} has however its limitations, since it includes either a forward or a reverse compression wave, but not both.  
Therefore, in their model, only IMF lines of the slow (fast) solar wind can intersect the  forward (reverse) compression wave, and as a result, the inner structure of the CIR is relatively simple, resembling a compressed Parker spiral magnetic field.  
However, in reality, there will often be both a forward and reverse  compression or shock wave bounding the CIR. 
In between those waves, magnetic field lines can converge to a stream interface (SI), which separates the compressed fast and slow solar wind plasmas, and which is expected to have a non-negligible influence on energetic particle transport \citep[see e.g.][]{intriligator01}.  
Such a CIR-structure is captured by our MHD simulation, and we show that it can have significant effects on the time intensity and anisotropy profiles of SEP events. 
In addition to the above, we  study how the magnetic field configuration inside the CIR can amplify the efficiency of cross-field diffusion, without requiring high levels of turbulence.

The structure of the article is as follows. 
In Section~\ref{sec:paricle_transport} we introduce the equations that our particle transport model uses to describe the pitch-angle dependent transport of energetic particles in the heliosphere.
Subsequently, in Section~\ref{sec:the_code} we describe the numerical details of the transport code, and its coupling to EUHFORIA. 
As a verification test, Section~\ref{sec:parker_seps} presents the performance of the model when describing the transport of protons under nominal slow and fast solar wind conditions. 
The effects of a different solar wind speeds on the particle distributions are discussed. 
In Section~\ref{sec:mixed_solar_wind} we present the results when propagating protons in a solar wind containing a CIR with different cross-field diffusion conditions. 
Finally, in Section~\ref{sec:summary} we summarise the results presented in this work and give the conclusions.
\section{Three-dimensional solar energetic particle transport model}\label{sec:paricle_transport}
The evolution of the gyrotropic phase-space distribution function $f(\vec{x},p,\mu,t)$ is described by the FTE, which can be written as \citep[e.g.][]{roelof69, isenberg97, zhang09, leRoux09}
\begin{equation}\label{eq:FTE}
\begin{aligned}
\pd{f}{t} &+\td{\vec{x}}{t}\cdot\nabla{f}+\td{\mu}{t}\pd{f}{\mu} +\td{p}{t} \pd{f}{p} \\
&= \pd{}{\mu}\left(D_{\mu\mu}\pd{f}{\mu}\right) + \nabla\cdot\left(\bm{\kappa}_\perp\cdot\nabla f\right),
\end{aligned}
\end{equation}
with
\begin{eqnarray}
\td{\vec{x}}{t} &= &
\vec{V}_{\rm sw}+ \vec{V}_d+\mu \varv\vec{b} \label{eq:fte_spat}\\
\td{\mu}{t}&=&
\frac{1-\mu^2}{2}\Bigg(\varv \nabla\cdot\vec{b} + \mu \nabla\cdot\vec{\vec{V}_{\rm sw}} - 3 \mu \vec{b}\vec{b}:\nabla\vec{\vec{V}_{\rm sw}}\label{eq:fte_mu}\\
\notag 
&&- \frac{2}{\varv}\vec{b}\cdot\td{\vec{\vec{V}_{\rm sw}}}{t} \Bigg) \\
\td{p}{t} &=&
 \Bigg( \frac{1-3\mu^2}{2}(\vec{b}\vec{b}:\nabla\vec{\vec{V}_{\rm sw}}) - \frac{1-\mu^2}{2}\nabla\cdot\vec{\vec{V}_{\rm sw}}\label{eq:fte_p} \\
\notag 
 && -\frac{\mu }{\varv}\vec{b}\cdot\td{\vec{\vec{V}_{\rm sw}}}{t}\Bigg) p.
\end{eqnarray}
Here $\vec{x}$ is the  spatial coordinate, 
${p}$ the momentum, 
$\varv$  the speed,
$\mu$ the pitch-angle cosine, 
$\vec{V}_{sw}$ is the solar wind velocity, and
$\vec{b}$ the unit vector in the direction of the mean magnetic field.
$D_{\mu\mu}$ is the pitch-angle diffusion coefficient, 
$\bm{\kappa}_\perp$ the spatial cross-field diffusion tensor, and
$\vec{V}_d$ the particle drift due to the gradient and curvature of the mean magnetic field.

The FTE is written in mixed coordinates, which means that the spatial coordinate is measured in the fixed inertial frame of the observer, whereas the momentum and pitch-angle cosine are  measured in a reference frame co-moving with the solar wind \citep[e.g.][and references therein]{zhang09}. 
In the latter frame,  the average electric field equals zero, leaving the Lorentz force only capable of changing the propagation direction of the particle and not its energy.  
However, we note that the co-moving frame is not inertial, and therefore fictitious forces  will act on the particle, altering its energy and pitch-angle.
 This results in a monotonic decrease of the momentum of the  particle in the co-moving frame when travelling in the expanding solar wind.
 We note that this is true even in the absence of particle scattering caused by  magnetic turbulence.
 This  constant loss of energy is often termed adiabatic deceleration, yet  if the particle travels in a converging flow, for example at shocks or compression regions,  the particle will accelerate instead of decelerate.  

In addition to adiabatic deceleration, \cite{dalla15} have shown that magnetic gradient and curvature drifts can also lead to a substantial deceleration of SEPs, especially at high energies and high latitude.
 These drift-induced energy losses  are included in the momentum terms Eq.~\eqref{eq:fte_mu} and Eq.~\eqref{eq:fte_p} of the FTE, as shown by \cite{leRoux09}. 
However, the standard FTE   normally does not include any effects due to magnetic drifts in the spatial convection term Eq.~\eqref{eq:fte_spat}. 
By extending this spatial term to include the missing drifts, one obtains the standard-drift guiding centre equation \citep{leRoux09}.  
This extension of the FTE is done in Eq.~\eqref{eq:FTE}, where  the effects of gradient/curvature drifts are included in the spatial part of the equation through 
\begin{equation}
\vec{V}_d = \frac{\varv p}{QB}\left[\frac{1-\mu^2}{2} \left( (\nabla\times \vec{b})_\parallel + \frac{\vec{b}\times\nabla B}{B}\right) + \mu^2 (\nabla\times \vec{b})_\perp \right],
\end{equation}
where $Q$ is the particle charge, and the subscripts $\parallel$ and $\perp$ denote, respectively, the parallel and perpendicular components with respect to the magnetic field $\vec{B}$.
We note that with this extension, the spatial coordinate $\vec{x}$ in the FTE represents the coordinates of the guiding center of a particle.  
In the above description, the effects of the polarization drift are not taken into account; \cite{dalla13} illustrated that this drift is significantly smaller than the sum of the gradient and curvature drifts. 

On their journey through the heliosphere, particles are subjected to electromagnetic forces resulting from the turbulence present in the solar wind. 
Since compressible wave modes are readily damped out in the solar wind,  Alfv\'en waves are typically considered as one of the main contributors to the solar wind turbulence \citep{Howes13}. 
To first order, the main effect of these waves is to scatter energetic particles elastically in the reference frame moving with the wave.  
Since the Alfv\'en speed is small compared to the speed of the solar wind, one can assume that the wave and solar wind frames coincide, meaning that the magnetic fluctuations are convected with the solar wind. 
Quasi-linear theory (QLT) then allows the description of the effects of magnetic slab turbulence by means of a diffusion process in pitch-angle space, with a diffusion coefficient given by \citep{jokipii66,jaekel92}
\begin{equation}
 D_{\mu\mu} = \frac{\pi}{2}\frac{1-\mu^2}{\varv |\mu|}\left(\frac{\Omega}{B}\right)^2P\left(k=\frac{\Omega}{|\mu| \varv}\right),
\end{equation}
where $\Omega$ denotes the gyrofrequency of the  particle, $k$ the parallel wave number, and $P$ the power spectrum of the magnetic turbulence. 
Assuming a power-law for the latter, that is, $P= Ck^{-q}$ with $C$ a proportionality constant, the diffusion coefficient adopts the form 
\begin{equation}\label{eq:D_std}
{D_{\mu\mu} = \frac{C\pi}{2} \left(\frac{|Q|}{m}\right)^{2-q}B^{-q}\varv^{q-1}\left(|\mu|^{q-1} +H\right)\left(1-\mu^2\right)},
\end{equation}
where  $m$ denotes the particle mass and the parameter $H$ is  added to describe the scattering through $\mu =0$ \citep[see, e.g.][]{beeck86}. 
When using this form of the diffusion coefficient in numerical applications, caution is needed since $\partial D_{\mu\mu} /\partial\mu$ has a pole at $\mu = 0$. 
This pole needs to be treated carefully in order to obtain correct pitch-angle distributions (PADs). In order to overcome any potential numerical issues caused by this pole, we follow the approach adopted by \cite{agueda08} and assume the following form for the pitch-angle diffusion coefficient
\begin{equation}\label{eq:D_neus}
{D_{\mu\mu} = \frac{\nu_0}{2} \left(\frac{|Q|}{m}\right)^{2-q}B^{-q}\varv^{q-1}\left(\frac{|\mu|}{1 + |\mu|} + \epsilon\right)\left(1-\mu^2\right)},
\end{equation}
which has the advantage that $\partial D_{\mu\mu} /\partial\mu$ is bounded for $\mu \in [-1,1]$. \cite{agueda13} illustrated that the diffusion coefficients defined in  \eqref{eq:D_std} and \eqref{eq:D_neus} closely match when the parameter $\epsilon$ is chosen carefully. 
The functional form \eqref{eq:D_neus} for the pitch-angle diffusion coefficient is in particular interesting for  Monte Carlo simulations, since it allows an  efficient implementation as explained in the following section and in \cite{agueda08}.

In addition to pitch-angle diffusion, Eq.~\eqref{eq:FTE} contains a cross-field spatial diffusion process described by the diffusion tensor  $\bm{\kappa}_\perp$. This term is not part of the standard FTE, yet is often added to the equation to describe the motion of particles perpendicular to mean magnetic field lines \citep[see, e.g.][]{zhang09,droge10,strauss15}. 
This diffusion tensor is typically chosen as 
\begin{equation}\label{eq:perp}
\bm{\kappa}_\perp = \kappa_\perp\left(\mathbb{I}-\vec{bb}\right),
\end{equation}
where $\mathbb{I}$ is the unit tensor and $\vec{bb}$ is a dyadic product.
In a reference frame aligned with the magnetic field, $\bm{\kappa}_\perp$  becomes then a diagonal matrix with only two non-zero elements.  
The perpendicular diffusion coefficient $ \kappa_\perp $ can be prescribed by using for example the non-linear guiding centre theory \citep{matthaeus03,shalchi10}.
 We note that the functional form \eqref{eq:perp} for the perpendicular diffusion tensor cannot  describe perpendicular motions due to particle drifts induced by turbulence, since this would require $\bm{\kappa}_\perp$  to have an anti-symmetric part.
 
Similarly to \cite{droge10}, we assume a perpendicular mean free path which scales with the gyro-radius of the particle. Averaging over the pitch-angle gives then the following diffusion coefficient
 \begin{equation}\label{eq:kappa_perp}
{ \bm{\kappa}_\perp = \frac{\pi }{12}\frac{\alpha\varv \lambda^r_\parallel}{ b^2_r}\frac{B_0}{B}}, 
 \end{equation}
where $\lambda^r_\parallel$ is the proton radial mean free path and $\alpha$ is a free parameter that determines the ratio between the parallel and perpendicular mean free path at a reference magnetic field strength $B_0$. Denoting the angle between the radial and the IMF direction by $\psi$, the radial mean free path is defined as $\lambda^r_{\parallel} = \lambda_{\parallel} \cos^2{\psi}=\lambda_{\parallel}b_r^2$. Moreover, the radial mean free path is related to the diffusion coefficient $D_{\mu\mu}$ through \citep{hasselmann70}
 \begin{equation}
\lambda_\parallel^r = \frac{3\varv b_r^2}{8}\int_{-1}^{1}\frac{\left( 1 - \mu^2\right)^2}{D_{\mu\mu}}d\mu,
\end{equation}
and is often assumed to be constant \citep[e.g.][]{bieber94}. This allows one to fix the proportionality constants $C$ and $\nu_0$ in Eqs.~\eqref{eq:D_std} and~\eqref{eq:D_neus}, respectively. Similar to \cite{droge10}, we will throughout this work assume  a constant radial mean free path of 0.3 AU for 4 MeV protons.  


\section{Numerical aspects of the model}\label{sec:the_code}

\subsection{Particle transport}

The FTE is a five-dimensional parabolic partial differential equation that can be solved using for example a finite difference method \citep[see, e.g.][]{ruffolo95,lario98,schwadron10,leRoux12,wang12} or by taking a stochastic approach \citep[see, e.g.][]{vainio98,agueda08,zhang09,droge10,strauss17}.
We take the latter approach and solve the FTE by means of a time-forward Monte Carlo simulation.
 In this section, we explain the numerical methods employed to propagate particles in phase space.

The time-forward stochastic differential equation (SDE) describing the spatial part of Eq.~\eqref{eq:FTE} is given by 
 \begin{equation}\label{eq:sde_spatial}
 d\vec{x} = \left(\td{\vec{x}}{t} +\nabla\cdot\bm{\kappa}_\perp \right)dt +\vec{A}\cdot d\vec{w}_\vec{x},
 \end{equation}
 where  $\vec{w}_\vec{x}$ is a Wiener process, and $\vec{A}$ is a $3\times 3$ matrix that satisfies $\vec{A}\vec{A}^ T =  2\bm{\kappa}_\perp$ \citep{gardiner04,pei10,strauss17}. To integrate Eq.~\eqref{eq:sde_spatial} we use the standard Euler–Maruyama method. 
   
To update the variable $\mu$, the standard SDE approach would be to use the Euler-Maruyama method  to integrate 
\begin{equation}
d\mu =\left(\td{\mu}{t} +\pd{D_{\mu\mu}}{\mu} \right)dt + \sqrt{2D_{\mu\mu}}dw_\mu,
\end{equation}
where $dw_\mu$ is a Wiener process. 
The behaviour of $\partial{D_{\mu\mu}}/\partial{\mu}$  near $\mu = 0$ requires the use of a very small time step to obtain correct PADs for values of $\mu$ near zero. In order to circumvent this and to gain computational efficiency, we instead apply the same method as \cite{agueda08} to update the cosine of the pitch-angle. In this approach, we integrate Eq~\eqref{eq:fte_mu}, which only contains deterministic terms, forward in time using the standard Euler method. To model the stochastic pitch-angle variations, we note that the diffusion coefficient, \eqref{eq:D_neus}, consists of an  isotropic pitch-angle scattering process described by
\begin{equation}
D^{\rm iso}_{\mu\mu} = \epsilon \frac{\nu}{2}\left(1-\mu^2\right), 
\end{equation}
and an  anisotropic pitch-angle scattering process described by
\begin{equation}
D^{\rm aniso}_{\mu\mu} = \frac{\nu}{2}\left(1-\mu^2\right)\left(\frac{|\mu|}{1 + |\mu|} \right),
\end{equation}
where in both cases $\nu = \nu_0\left({|Q|}/{m}\right)^{2-q}B^{-q}\varv^{q-1}$.
The  anisotropic scattering process can be made isotropic by performing a coordinate transformation; thus, pitch-angle scattering in our model is treated by means of two isotropic scattering processes \citep[see details in][]{agueda08}.
For an isotropic scattering process with scattering frequency $\nu$, the  pitch-angle distribution around the propagation direction of the unscattered particle is, after a certain time $\delta t \ll \nu^{-1}$, given by \citep{torsti96,vainio98}
\begin{equation}\label{eq:scatter_distr}
F(\vartheta,\varphi,\delta t)d\Omega = \frac{1}{2\pi}\left[\frac{1}{2 \nu \delta t} \exp\left(-\frac{\vartheta^2}{2 \nu \delta t} \right)d\vartheta^2\right]d\varphi,
\end{equation}
where $\vartheta$ is the angle between propagation directions of the unscattered and scattered particle trajectory, and $\varphi$ is the phase angle around the scattering axis. Following Eq.~\eqref{eq:scatter_distr}, the new  cosine of the pitch-angle,  $\mu_{\rm new}$, is then related to the pitch-angle cosine $\mu_{\rm old}$ from before the scattering process through \citep{vainio98}
\begin{equation}
\mu_{\rm new} = \mu_{\rm old}\cos\vartheta + \sqrt{1-\mu_{\rm old}^2}\sin\vartheta\cos\varphi.
\end{equation}

Finally, the magnitude of the particle's momentum is updated by integrating  Eq.~\eqref{eq:fte_p} forward in time using the Euler method.

By solving the FTE in a time-forward manner, we obtain the particle differential flux $j(\vec{x},p,\mu)$, defined here as the number density of particles in phase space element $2 \pi d\vec{x}dpd\mu$, and related to the particle distribution function $f(\vec{x},p,\mu)$ through $j =  p^2f$. In practice, we calculate $j(t,\vec{x},E,\mu)$ by sampling the simulated particles in five-dimensional volume elements defined as $(\vec{x}+\Delta\vec{x},E+\Delta E,\mu+\Delta\mu)$, where $E=p^2/(2m)$ represents the kinetic energy of the particle, and $m_p$ its mass. To increase the statistics, we average over a time period $t + \Delta t$ to obtain a representation of   $j(t,\vec{x},E,\mu)$. Representing the time step of the particle code by $\delta t$, and choosing $\Delta t = n \delta t$, with $n \in \mathbb{N}$, we obtain  the differential flux thus as 
\begin{equation}
\frac{1}{\Delta t} \int_{t}^{t+\Delta t} \frac{ dN}{2 \pi d\vec{x}d\mu dE/\varv}dt \approx 
\frac{1}{\Delta t}\sum_{k =0}^{n} \frac{\varv \Delta N_k}{2 \pi \Delta\vec{x}\Delta E\Delta\mu}{\delta t}
\end{equation}
where $\Delta N_k$ gives the number of particles in a five-dimensional sampling volume at time $t + k\delta t$.  For the results presented in this work, we chose $\Delta\vec{x} =(0.025 \rm{\,AU})^3 $, $d\mu = 0.1$, $\Delta E = 0.2$ MeV, and $\Delta t = 10$ minutes. 
The sampling volumes cover the entire region of interest in phase space, allowing  a continuous coverage in energy and pitch-angle space at any location in the heliosphere. Finally we note that the code has reflective inner and absorptive outer boundary conditions. The outer boundary is placed at sufficiently large radial distance such that it does not affect the simulation results.

\subsection{Coupling with EUHFORIA}

In order to describe the transport of particles in the 3D heliosphere, a model is needed for the description of the IP  medium in which particles are accelerated and transported. 
The SEP transport model presented in this work is able to propagate particles in a solar wind generated by the EUHFORIA model \citep{pomoell18}. 
EUHFORIA consists of a data-driven coronal model and an MHD heliospheric model. The coronal part can use observations of the photospheric magnetic field to construct a model of the coronal large-scale magnetic field which, in turn, is used to obtain solar wind plasma parameters at 0.1 AU, by employing a set of empirical relations. 
Subsequently, the heliospheric model utilises these plasma parameters as boundary conditions to solve the ideal MHD equations augmented with gravity up  to a prescribed outer boundary, which can be several astronomical units. We have set this outer boundary to 5 AU  in order not to introduce any artificial
effects due to a limited domain on the particle  distributions obtained.
 Although it is not used in this work, EUHFORIA also allows the ejection of CMEs in the ambient solar wind of the heliospheric model. For further details about the implementation of EUHFORIA, we refer to \cite{pomoell18}. 

We note that the particle transport model propagates particles in a grid-free numerical scheme, whereas EUHFORIA provides the solar wind plasma variables on a numerical grid.
 Therefore, to integrate the particle transport equations, we interpolate the EUHFORIA variables to the location of the particle at each time step by means of tri-linear interpolation. 
 The solar wind configuration used in this work is stationary in the corotating frame (see Section~\ref{sec:mixed_solar_wind}).  Therefore, by performing a suitable coordinate rotation, the MHD and particle time steps coincide.


\section{ Propagation of solar energetic particles in uniform wind conditions}\label{sec:parker_seps}
    \begin{figure}
        \centering
        \includegraphics[width=0.4\textwidth]{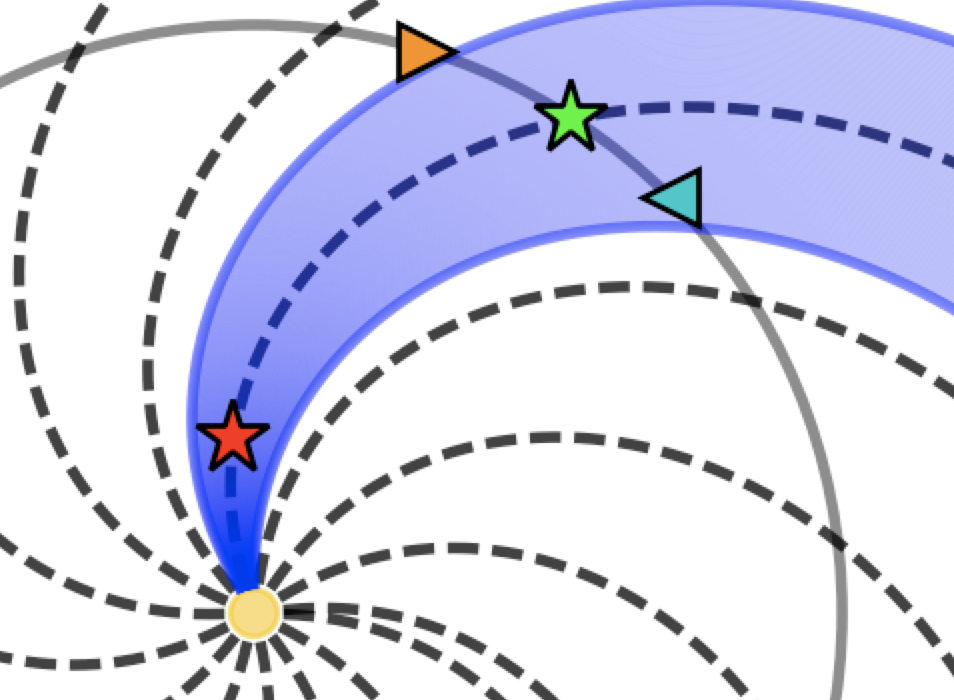}
    \caption{Diagram of the ecliptic illustrating the particle streaming zone (blue) for an injection over an azimuthal range of $30^\circ$. The markers represent the different observers introduced in the text. The triangles are stationary observers A (cyan) and B (orange), whereas the stars are the corotating observers C (red) and D (green). The dashed lines represent magnetic field lines. }
     \label{fig:diagram}
\end{figure}

    \begin{figure*}
        \centering
        \begin{tabular}{cc}
        \includegraphics[width=0.4\textwidth]{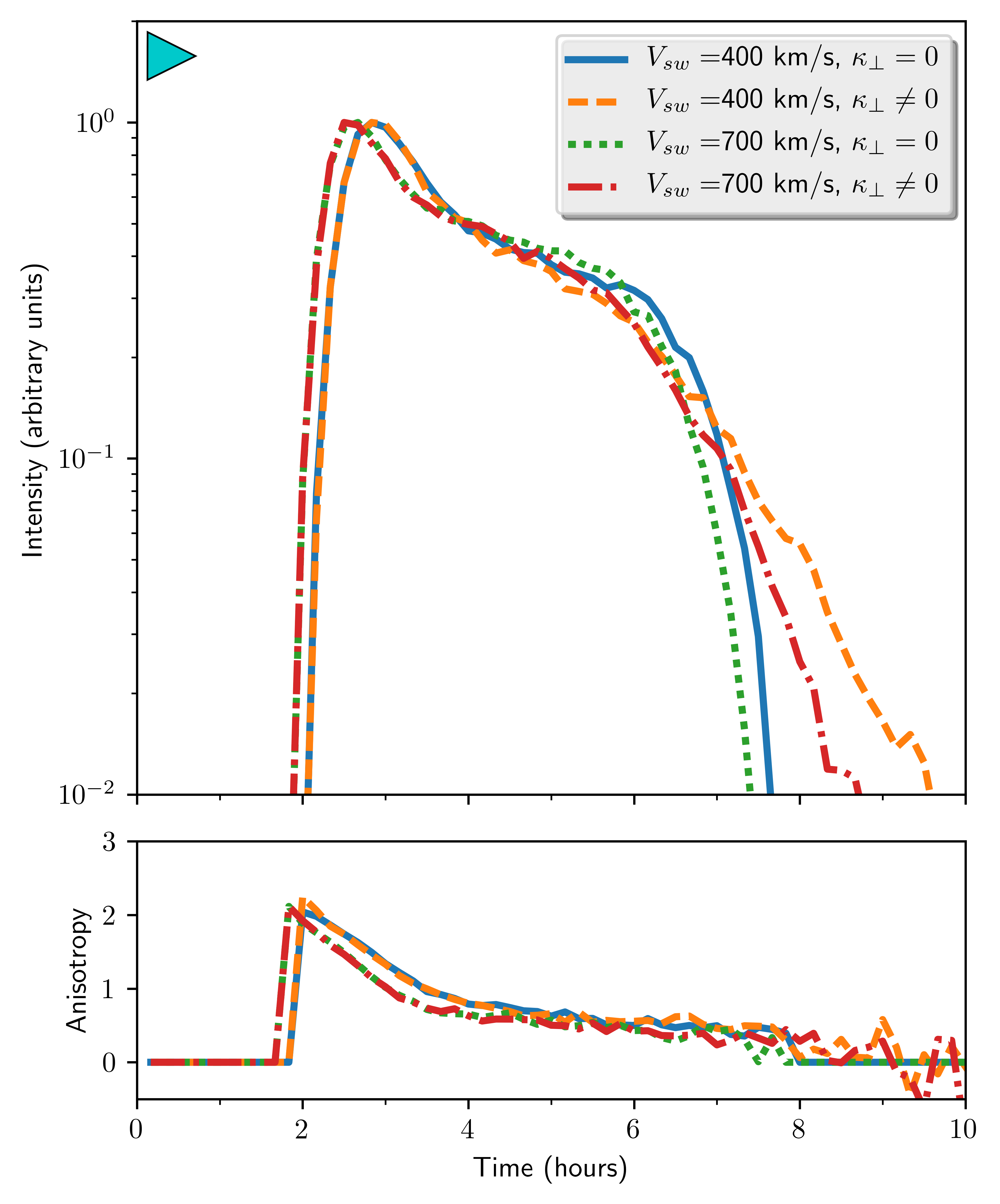}&
        \includegraphics[width=0.4\textwidth]{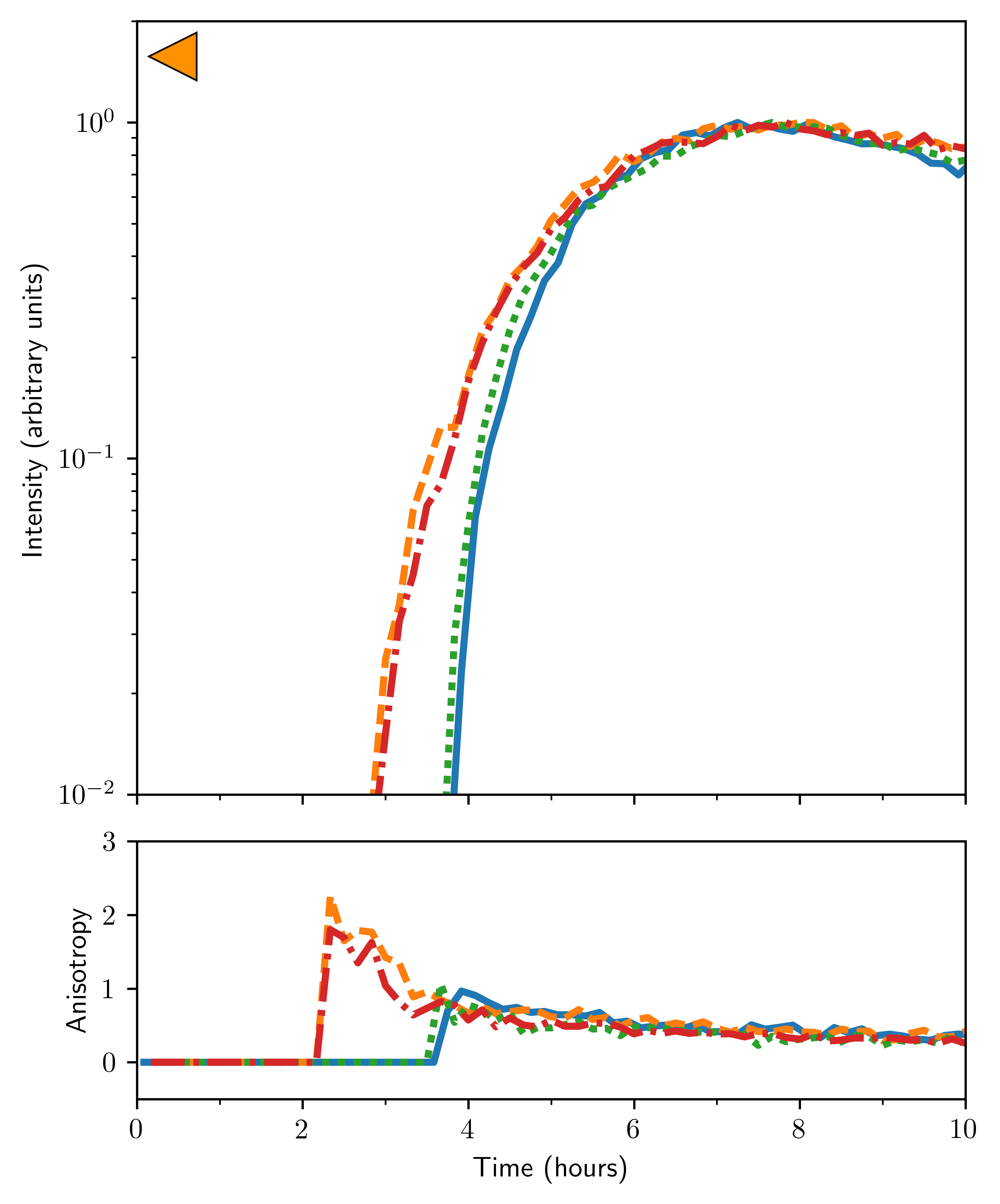}\\
    \end{tabular}
    \caption{Intensity and anisotropy time profiles of $4$ MeV protons propagating in fast and slow solar wind configurations, with and without cross-field diffusion as detailed in the legend. The left (right) panel shows the results for observer A (B) indicated by the cyan (orange) triangle marker as in Fig.~\ref{fig:diagram}. Adiabatic deceleration and solar wind convection are neglected, with the exception of accounting for corotation.}
     \label{fig:noDecParker}
\end{figure*}
As a verification test, we present in this section the results obtained when propagating particles in a nominal IMF for slow ($400$ km/s) and fast ($700$ km/s) solar wind configurations. 
Both solar winds are computed assuming a constant sidereal solar rotation period of  $25.4$ days. Similar to \cite{droge10}, we consider an impulsive event of $4$ MeV protons, injected uniformly at the solar equator over a longitudinal range of $30$ degrees and at a radial distance of $0.05$~AU. 
For each simulation presented in this section, we  propagated $10^6$ particles.

As illustrated in the diagram shown in Fig.~\ref{fig:diagram}, we place four different observers  in the solar equatorial plane.  
The two observers marked by triangles are stationary. Observer~A (cyan triangle) is initially located in the SEP streaming zone, close to the eastern boundary, while observer~B (orange triangle) is initially located outside the SEP streaming zone, near the western boundary.
 Eventually, the corotation of the particles with the magnetic field leaves observer~A outside the SEP streaming zone, whereas  observer~B  moves into the SEP streaming zone. 
 The two remaining observers, C (red star) and D (green star), are corotating with the centre of the particle streaming zone and located at heliocentric radial distances of $0.3$~AU and $1$~AU, respectively.

 As a first test, we transport particles  considering neither energy losses due to adiabatic deceleration nor particle drifts. 
Therefore, the particles retain their original $4$ MeV injection energy throughout the entire simulation. We also neglect solar wind convection, but include the corotation effect.
 Similar assumptions were made by \cite{droge10}, and their results are used here to verify the results of the transport code.
 With this aim, we assume $\epsilon =0.048$ in Eq.~\ref{eq:D_neus}, to characterise $D_{\mu\mu}$.
 This value was obtained from minimizing the sum of the squared differences between  \eqref{eq:D_neus} and \eqref{eq:D_std} for $H=0.05$, as this latter value was used by \cite{droge10}. 
 We assume a constant radial mean free path, $\lambda_\parallel^r =0.3$ AU, and perform simulations both with and without spatial cross-field diffusion. 
 For the cases with cross-field diffusion, we assume $\alpha=10^{-4}$ in Eq.~\eqref{eq:perp}, which corresponds to a relatively small perpendicular mean free path. We note that for example \cite{droge16}, \cite{qin15}, and \cite{strauss17b} use a $\kappa_{\perp}$ of the order of $\sim 0.01\kappa_{\parallel}$, that is, about $100$ times larger than the value we assume. 

Figure~\ref{fig:noDecParker} shows the intensity, $ I = \frac{1}{2}\int_{-1}^{1} j(\mu)d\mu$, and the anisotropy, $A = 3\int_{-1}^{1} \mu f(\mu)d\mu / \int_{-1}^{1} f(\mu)d\mu$, measured by  the two stationary observers described above. 
The intensity profiles depicted in the left panel of this figure (observer~A)  for the cases where particles travel in the slow solar wind (blue and orange curves) reproduce the profiles shown in the right panel of Fig.~8 of \cite{droge10}.
 The effects of cross-field diffusion are seen  when observers~A and B leave and enter, respectively, the particle streaming zone. For observer~A and for both solar winds, the cut-off in the particle intensity observed after ${\sim} 7$~hours becomes more gradual due to cross-field diffusion.
For observer B (right panel of Fig.~\ref{fig:noDecParker}), the effect of cross-field diffusion is to shift the particle onset to an earlier time.
    \begin{figure*}
        \centering
        \begin{tabular}{cc}
            \includegraphics[width=0.4\textwidth]{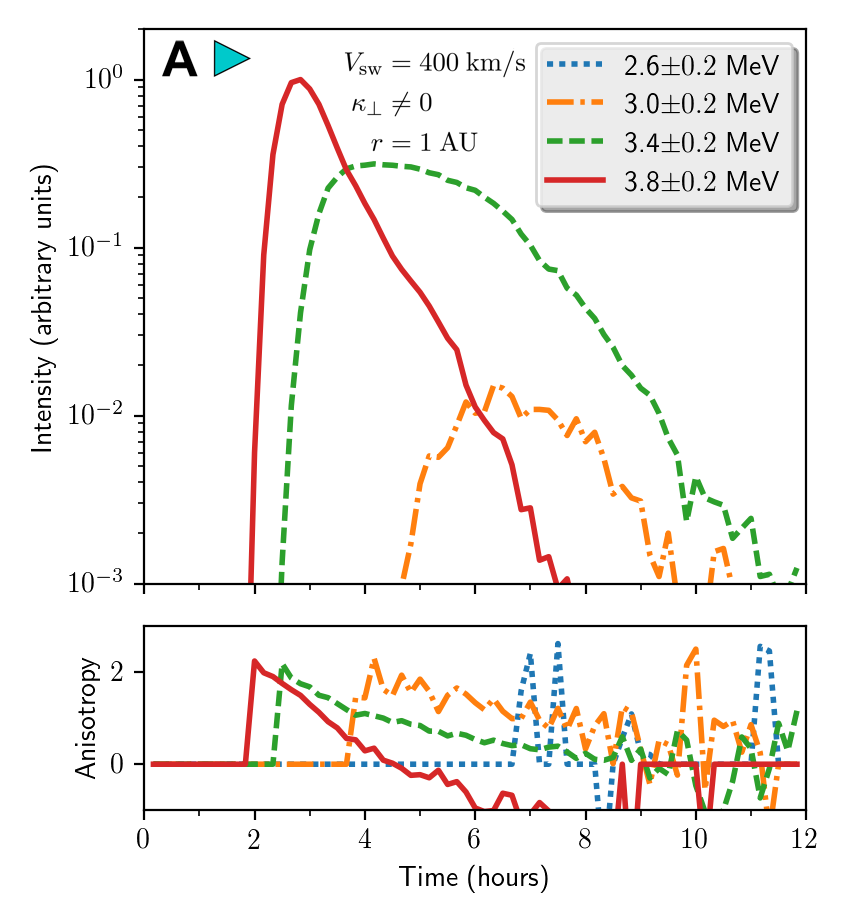} &
            \includegraphics[width=0.4\textwidth]{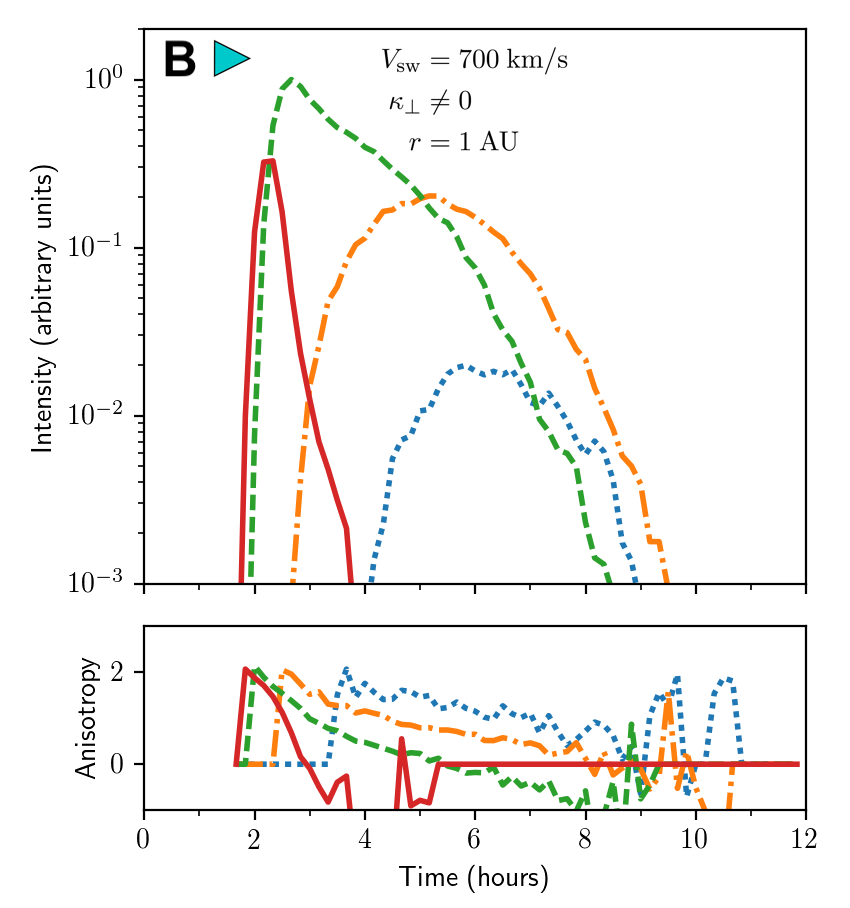}  \\
            \includegraphics[width=0.4\textwidth]{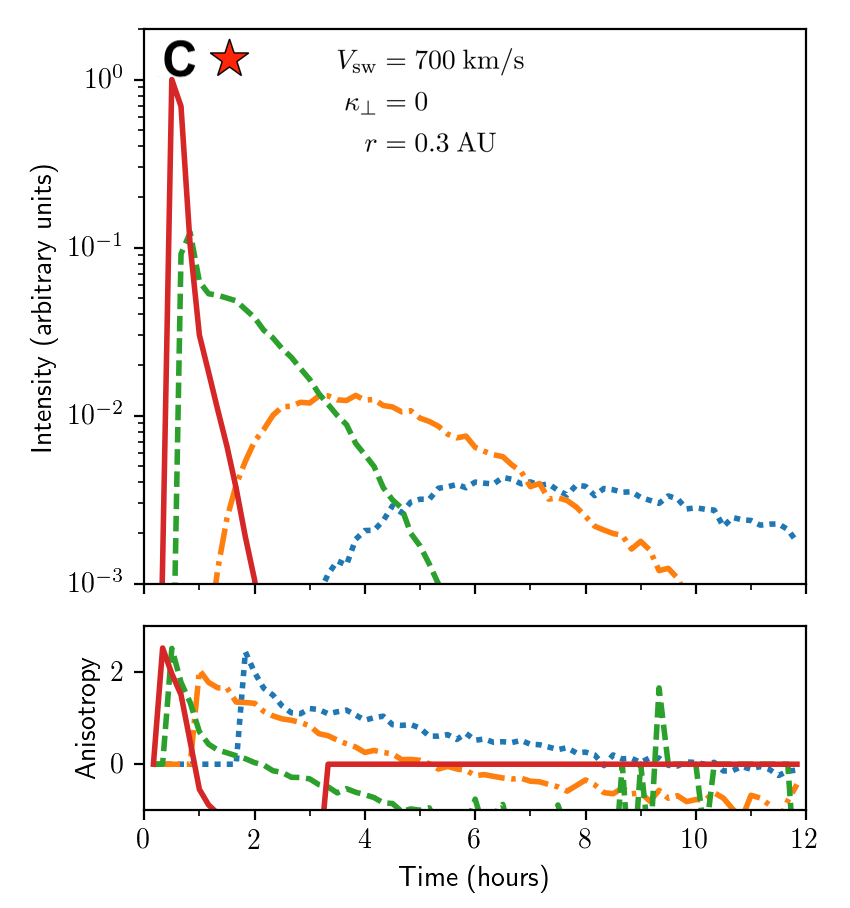} &  
            \includegraphics[width=0.4\textwidth]{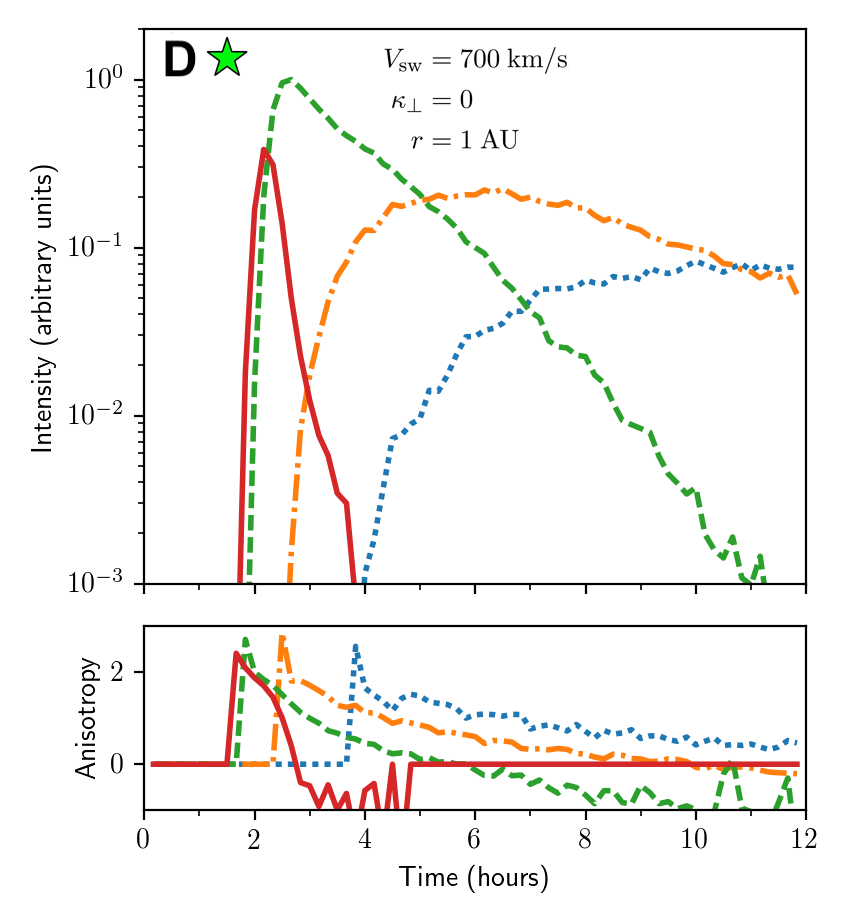}\\
        \end{tabular}
        \caption {Intensity and anisotropy time profiles for different energy channels of protons injected with an initial energy of $4$ MeV. All the effects of the FTE are included. The first row shows the profiles for observer~A located in the slow (panel~A) and fast (panel~B) solar wind. The last row shows the profiles for corotating observers~C (panel~C) and D (panel~D).}
        \label{fig:DecParker}
    \end{figure*}

The effect of the different solar wind speeds is relatively weak for these simulations and merely reflects the difference in the IMF curvature of the solar wind configuration,  that is, the distance a particle needs to travel to reach $1$~AU reduces for larger values of the solar wind speed. 

In contrast, the effect of the solar wind speed becomes much more pronounced when all terms of the FTE are included in the simulations.
 Particle intensity-time profiles and anisotropies obtained from such simulations are shown in Fig.~\ref{fig:DecParker} for the same observers and different energy channels, in the range of 2.2\,--\,4.0 MeV.
Panel~A of Fig.~\ref{fig:DecParker} shows the results for observer~A when located in the slow solar wind.  
Recalling that all particles were injected with the same initial energy of $4$ MeV, the effect of adiabatic deceleration becomes very clear. 
Only during the first one and a half hours after the particle onset time, is the main contribution to the total intensities due to particles from the highest energy channel,  which subsequently falls off rapidly by several orders of magnitude, leaving  $3.4 \pm 0.2$ MeV as the most populated energy channel.
 The latter remains the main contributor to the total intensity until the observer leaves the particle streaming zone, and all intensities drop to zero. 

Panel~B of Fig.~\ref{fig:DecParker} shows the particle intensities for observer~A, but this time for the fast solar wind case.
The comparison between panels~A and B shows that adiabatic deceleration is much stronger in the fast solar wind. In this latter case, the majority of particles have already decelerated out of the highest-energy channel upon reaching $1$ AU. 
At a time of$~5$~hours, the $3.0 \pm 0.2.$ MeV channel becomes the most populated energy channel, a transition that does not occur for the intensities measured in the slow solar wind case.  
In addition, a considerable number of particles are registered in the $2.6 \pm 0.2$ MeV channel in the fast solar wind case, whereas this channel is only scarcely populated in the slow solar wind simulation.
 The strong dependence of adiabatic deceleration on the solar wind speed can be understood by writing  Eq.~\eqref{eq:fte_p} for the case of a constant radial velocity, giving  $d{p}/d{t} \propto V_{sw}$ \citep[see, e.g.][]{ruffolo95}.

The bottom panels of Fig.~\ref{fig:DecParker} show the intensity and anisotropy time profiles for the two corotating observers. At 0.3~AU (panel~C), the event onset is characterised by a sharp peak of SEPs populating the highest-energy channel  (red curve). 
This peak reflects the impulsive mono-energetic delta injection, and corresponds to the particles that have not yet scattered significantly on their path from $0.05$ to $0.3$ AU. 
Moreover, we note that the intensity of the highest-energy channel shows a kink around \textasciitilde 1~hour, corresponding to the time from when the majority of the particles start coming from the anti-sunward direction, as indicated by the change of sign seen in the corresponding anisotropy time profile. 
The intensity and anisotropy profiles for the corotating observer located at $1$ AU (Panel D) show similar features to those in Panel~B, with the exception that after \textasciitilde 11 hours the lowest-energy channel depicted becomes the most populated one because the corotating observer remains in the particle streaming zone.
In addition to the effects of adiabatic deceleration, the qualitative differences between the intensity profiles for the corotating observers at 0.3~AU and 1.0~AU are partly due to the fact that particles with the highest energy have the hardest time travelling back in sunward direction due the magnetic mirroring effect, which is proportional to both the speed of the particle and the magnetic field magnitude. 
Therefore, at $0.3$ AU, the strong focusing/mirroring effect drives particles of higher energy more effectively towards large radial distances than particles of low energy. 
This explains why the transition from energy channel $3.0 \pm 0.2$ MeV being the most populated to $2.6 \pm 0.2.$ MeV being the most populated occurs earlier in panel~C than in panel~D.



\section{Propagation of solar energetic particles in mixed solar wind conditions}\label{sec:mixed_solar_wind}
\subsection{A corotating interaction region }\label{sec:cir}
  \begin{figure*}
        \centering
        \begin{tabular}{cc}
        \includegraphics[width=0.45\textwidth]{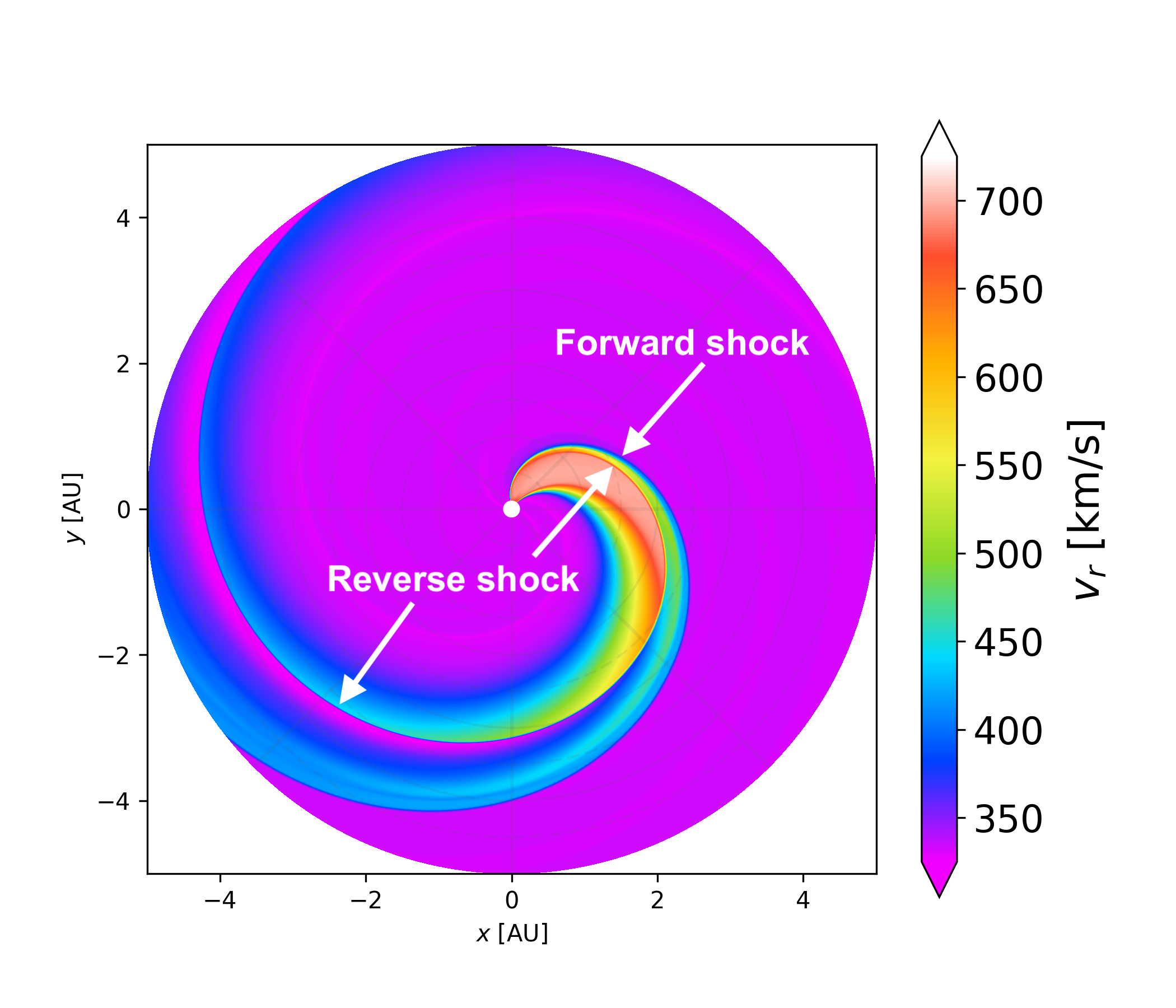}&
        \includegraphics[width=0.45\textwidth]{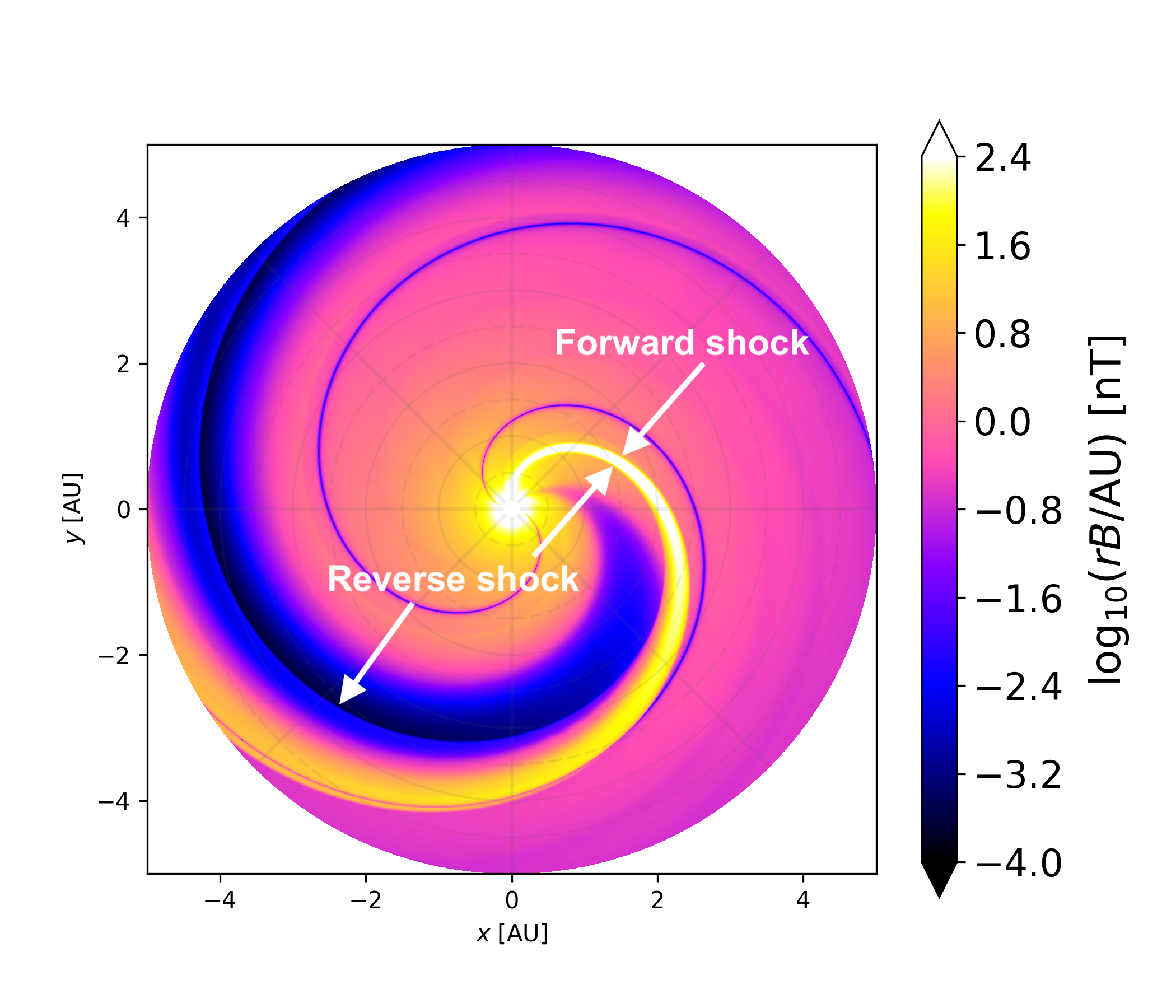}\\
    \end{tabular}
    \caption{Snapshot from the EUHFORIA simulation in the heliographic equatorial plane. The left panel shows the radial speed, while the right panel shows the logarithm of the scaled magnetic field magnitude, $rB$.}
     \label{fig:euh_color}
\end{figure*}
  \begin{figure}
        \centering
        \includegraphics[width=0.4\textwidth]{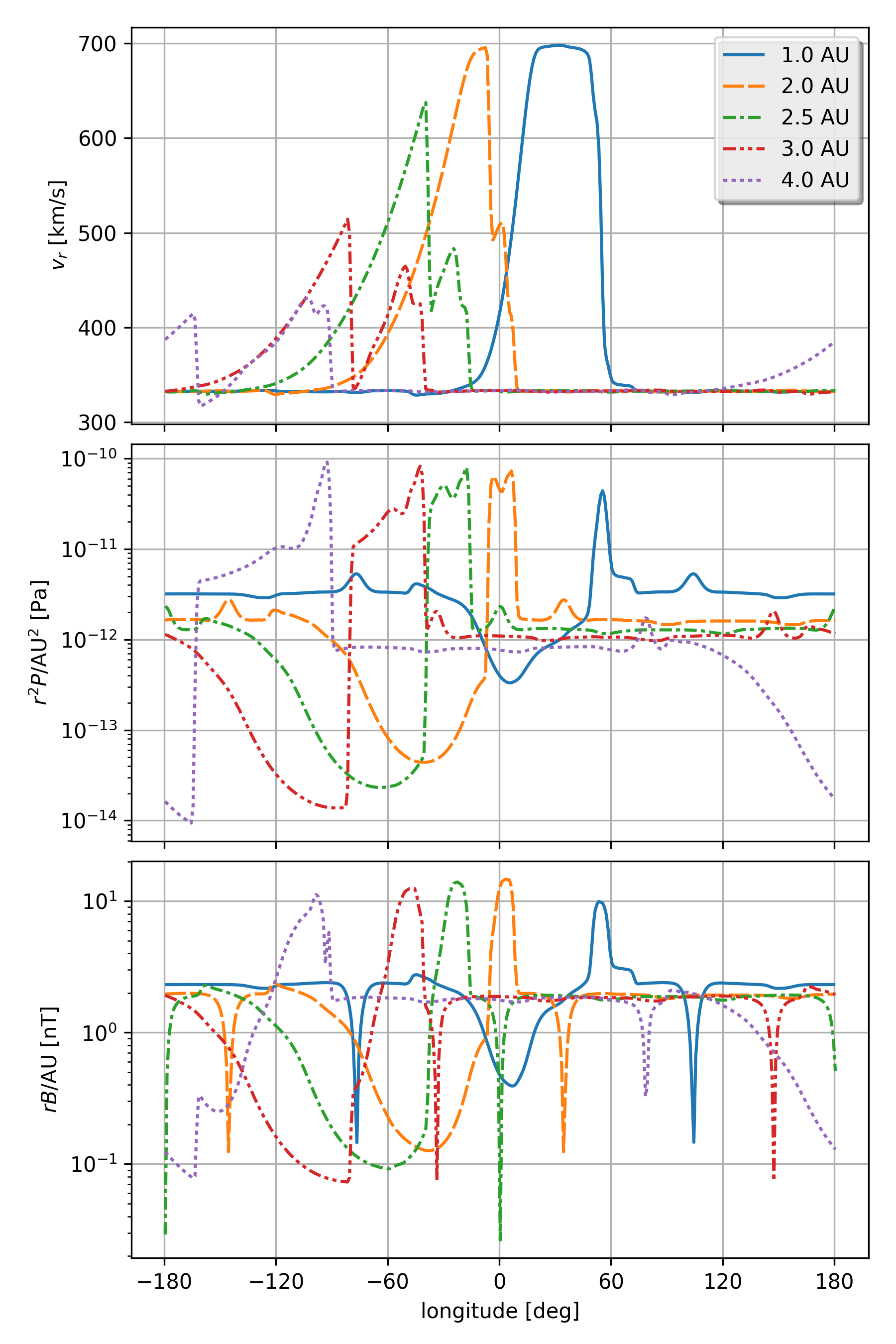}
    \caption{Longitudinal profiles of the radial velocity (top panel), scaled thermal pressure (middle panel), and scaled magnetic field magnitude (bottom panel) at different radial distances. }
     \label{fig:euh_lon}
\end{figure}

Now, we study the propagation of SEPs in a non-nominal solar wind configuration computed by EUHFORIA. We demonstrate the performance of the coupling of the SEP transport model with the EUHFORIA model by propagating particles in a synthetically generated solar wind. 
In particular, we simulate a fast solar wind stream embedded in a slow solar wind. 
Such a configuration is obtained by prescribing a solar wind with a speed of $330$~km~s$^{-1}$ everywhere at the inner boundary of the heliospheric model of EUHFORIA, except for the points with longitude and latitude satisfying 
\begin{equation}\label{eq:fsw}
{(\rm{longitude}-75^\circ)^2 + (\rm{latitude}-5^\circ)^2}< (20^\circ)^2,
\end{equation}
 where we prescribe a faster solar wind with a speed of $v_{\rm fsw} = 660$~km~s$^{-1}$. To some extent, this is reminiscent of a coronal hole located close to the solar equatorial plane. 
 The fast solar wind region described by \eqref{eq:fsw} is surrounded by a transition region 
 of an angular width of $6^\circ$, in which the speed changes linearly from the slow to the fast solar wind speed. EUHFORIA also requires the prescription of the polarity of the magnetic field at the inner boundary. We choose a dipolar polarity structure, with the current sheet tilted by $40^\circ$ with respect to the solar equatorial plane, intersecting the latter at longitudes $165^\circ$ and $345^\circ$. 
 In this configuration, the fast solar wind stream is completely embedded in a magnetic field of positive polarity. As described in \cite{pomoell18}, the number density $n$, and the magnetic field vector are prescribed at the inner boundary $R_b$, so that a constant kinetic energy density is obtained. This is done by choosing these variables as follows 
 \begin{eqnarray*}
 \left(Br, B_\theta, B_\phi\right) &=& \left({\rm sgn} (B_{\rm dp})B_{\rm fsw}v_r/v_{\rm fsw},0,-(B_r/v_r)R_b\Omega_\odot \sin\theta\right)\\
 n &=& n_{\rm fsw}(v_{\rm fsw}/v_r)^2,
\end{eqnarray*} 
 where  $n_{\rm fsw} = 300 \, \rm  cm^{-3}$  and $B_{\rm fsw} = 300 \, \rm  nT$ are  the number density and magnetic field strength of the fast solar wind, respectively, and ${\rm sgn} (B_{\rm dp})$ is the sign of the magnetic dipolar structure described above. Finally, the plasma thermal pressure on the inner boundary is chosen to be constant  and equal to $P = 3.3\, \rm nPa$. 

The EUHFORIA simulation is performed with a radial resolution of $1.03$\,R$_\odot$, and an angular resolution of $1^\circ$ for both longitude and colatitude, resulting in a numerical grid consisting of $1024\times120\times360$ cells. The simulation is started by performing a relaxation in which the MHD  equations are advanced in time until a fully steady-state solar wind  is obtained in the corotating frame. 
 
Figure~\ref{fig:euh_color} displays  snapshots of the solar wind simulation showing the radial velocity (left panel) and the scaled magnetic field magnitude, $rB$, (right panel), in the solar equatorial plane. 
Despite the simple inner boundary conditions, the generated solar wind contains a non-trivial structure. 
The substantial difference between the prescribed speeds of the slow and fast winds, combined with a relatively narrow transition region, results in the formation of a  CIR from relatively small radial distances outwards ($r > 1.1$~AU). 
This CIR is bounded by a forward shock wave moving into the slow solar wind and a reverse shock wave moving into the fast solar wind and  the rarefaction region behind the fast solar wind. 
We note that due to the finite resolution of the simulation, the width of the shock wave in the simulation is  larger than it would be for a real interplanetary CIR shock. 
This makes it difficult to exactly pinpoint where the boundaries of the CIR evolve from a large amplitude wave to a fully formed  shock wave. 
Hence, to estimate the location of the shock formation, we searched for 
strong jumps in radial and longitudinal profiles of various MHD quantities. 
As an example, Fig.~\ref{fig:euh_lon} shows, from top to bottom, the longitudinal profiles of the radial velocity, the scaled thermal pressure, and the scaled magnetic field magnitude at different radial distances (colour coded as indicated in the inset). 
By examining the top panel and following the curves in the direction of decreasing longitude, we see that at 1~AU (blue curve) the increase in the radial speed profile is still relatively smooth. From 2~AU outward, the increase splits into two steps, indicating the formation of the forward-reverse pair of shocks between 1 and 2 AU, that become increasingly clearly separated at larger radial distances. 
We note that the second increase in solar wind speed is simultaneously seen with a decrease in both the solar wind pressure and IMF profiles, as respectively shown in the middle and bottom panels of Fig.~\ref{fig:euh_lon}, indicating the presence of a reverse shock. 
Although at 1 AU the longitudinal profile of the radial velocity does not show a separation between the forward and the reverse shock, the pressure profile  does exhibit a clear two-step transition, indicating that the shocks are starting to form.

Both the forward and reverse shock locations become apparent in Fig.~\ref{fig:euh_color}. The forward shock wave is seen in the radial speed contours (left panel) as sudden colour jumps from purple to blue, and from pinkish to yellow in the scaled magnetic field (right panel). In this forward shock, the slow solar wind plasma is accelerated and compressed and the magnetic field magnitude increases significantly in the downstream region. We also  note that the current sheet, indicated as a dip in the magnetic field intensity, crosses the forward shock at a distance of ${\sim} 3$ {AU} (see the right panel of Fig.~\ref{fig:euh_color}). 

At small radial distances ($r<2.5$ AU), the reverse shock is observed as colour jumps from yellow to reddish orange in both left and right panels of  Fig.~\ref{fig:euh_color}. 
At larger radial distances, the reverse shock becomes much more clear as indicated by the sharp colour transitions from purple to blue (left panel) and from blue to black (right panel). 
The reverse shock decelerates and compresses the fast solar wind, and increases the magnetic field in the shocked plasma. 
We point out that, in our simulation, the reverse shock is in fact travelling in the rarefaction region behind the fast solar wind stream at  large radial distances. 
This is due to the relatively small size of the source region of the fast solar wind. 
 
 \begin{figure}
        \centering
        \includegraphics[width=0.5\textwidth]{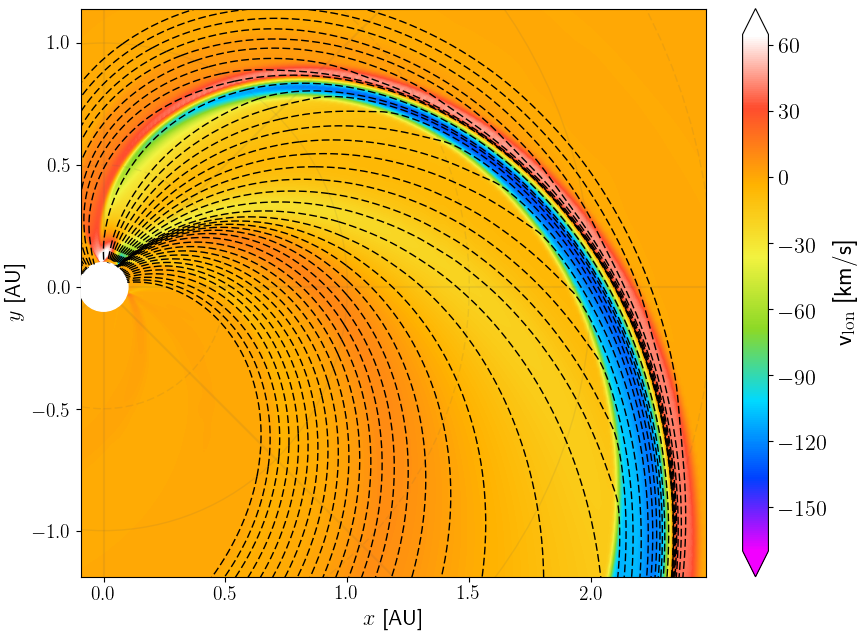}
    \caption{Longitudinal velocity profile of the solar wind in the heliographic equatorial plane, illustrating the streaming interface (the yellow stripe separating the blue and reddish bands). The black dashed lines are magnetic field lines, drawn with a constant longitudinal separation of $3.5^\circ$ at $r = 1$ AU. }
     \label{fig:SI}
\end{figure}  
As illustrated in Fig~\ref{fig:SI}, the SI is characterised by a reversal of the longitudinal solar wind flow angle {(see the yellow stripe separating the blue and reddish bands)}, since the shock waves and the interaction between the two shocked plasmas deflect the solar wind in different directions. 
Figure~\ref{fig:SI} also shows how the IMF converges towards the SI inside the CIR; thus embedding the SI in a strong magnetic field. In addition, although not shown here, the SI is characterised by a relatively abrupt change in plasma density. As discussed below, the different properties of the shocked fast and slow solar wind plasmas, and their SI, may have a non-negligible effect on the energy changes of energetic particles. 

\subsection{SEP transport in the heliosphere}
  \begin{figure*}
        \centering
        \includegraphics[width=0.9\textwidth]{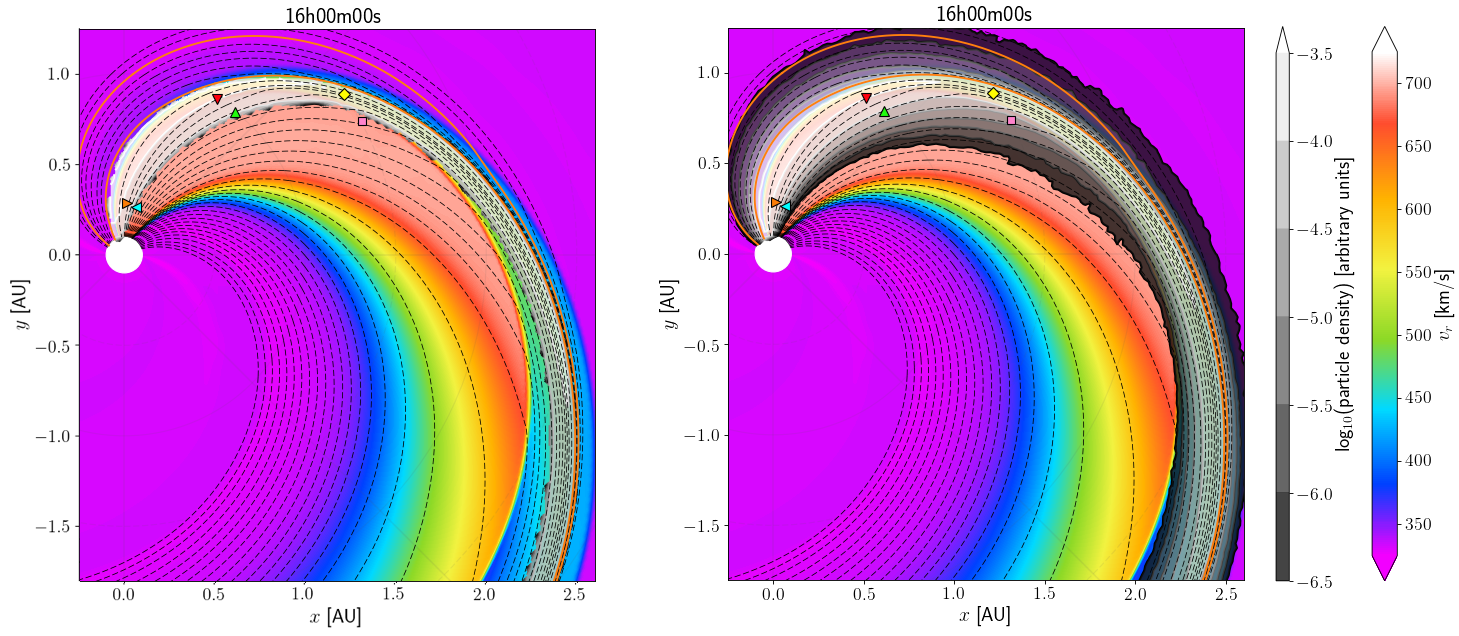}
    \caption{Particle densities drawn in grey shades on top of the radial velocity profile of the background solar wind, 16 hours after the particle injection (online movie). The left and right panels correspond to the cases without and with perpendicular diffusion, respectively. The solid orange  and dashed black  lines represent magnetic field lines, drawn with a constant longitudinal separation of $3.5^\circ$ at $r = 1$ AU. The markers correspond to the location of the different observers discussed in the text.}
     \label{fig:all_particles}
\end{figure*}

We consider a delta injection in time of $10^8$ protons with an energy of 4 MeV,  an isotropic pitch-angle distribution, and uniformly distributed over a spatial region described by
 \begin{equation}\label{eq:inj_region}
 (\rm{longitude},\rm{latitude})\in [75^\circ,105^\circ]\times [-5^\circ, 5^\circ]
 \end{equation}
at the inner boundary of the heliospheric model.
 Compared to the source of the fast solar wind (see Eq.~\eqref{eq:fsw}), the particle injection region covers a small part of the slow solar wind, the transition zone, and a significant part of the fast solar wind.
Such a configuration, in which energetic particles are impulsively injected near the boundary of a coronal hole, may especially be representative for $^3$He-rich SEP events \citep[see e.g.][]{wang06,kocharov08,bucik18}. 
  
The choice of our injection region also ensures that the particles do not interact with the current sheet during the 16~hours that we propagate the particles. The effects of the current sheet on the particle distributions near a CIR is left for future work.
 
 Using a mono-energetic particle injection allows us both to study the energy changes of the particles more easily and to compare the results with those obtained in Section~\ref{sec:parker_seps}.
As in Section~\ref{sec:parker_seps}, we assume a constant radial mean free path $\lambda_\parallel^r = 0.3$  AU for 4 MeV protons throughout the entire heliosphere. 
Since we are using a constant radial mean free path, the actual parallel mean free path will vary according to $\lambda_\parallel = \lambda_\parallel^r/b_r^ 2 $. At a fixed radial distance, $b_r$ is larger in the fast solar wind than in the slow solar wind, and hence we have a smaller parallel mean free path in the former as compared to the latter. In the CIR, the parallel mean free path will have values between those of the fast and slow solar winds.  
Observations have shown that the parallel mean free path in the fast solar wind is typically smaller than in the slow solar wind \citep[see, e.g.][]{erdos99}.
Nevertheless, we remark that  the mean free path of the  particles might vary rather strongly across solar wind regimes with different flow speed \citep[see, e.g.][]{pacheco17}, and hence our assumption of a constant  $\lambda_\parallel^r$ might underestimate this variation.

We perform simulations both with and without cross-field diffusion.  We note that the simulations without cross-field diffusion do include the effect of particle drifts, and hence there is  a possibility for the particle to move perpendicular to the IMF, although these drifts are very small for 4 MeV protons travelling at low latitudes. 
To characterise $\kappa_{\perp}$ we assume as in Section~\ref{sec:parker_seps} that $\alpha=10^{-4}$ in Eq.~\eqref{eq:kappa_perp}.  
Moreover, we choose the reference magnetic field magnitude $B_0$ to be the maximum magnetic field strength at 1 AU, that is, $B_0 = \max_{r=1\rm \, AU}B = 9.7 \, \rm nT$. 
This maximum  is obtained in the compressed shocked slow solar wind, and hence the cross-field diffusion is  the smallest there.
 In contrast, the shocked fast solar wind contains a magnetic field (${\sim}2\, \rm nT$ )  that is   significantly  smaller  than that of the shocked slow solar wind. 
 This means that the cross-field diffusion will be stronger in the shocked fast solar wind.
 The SI separates these  plasma populations and therefore also acts as a boundary across which almost no particle can diffuse in our simulations.
This is clearly seen in our simulations when injecting particles only in the fast solar wind regime (not shown here).
  Since our injection region is connected to magnetic field lines at both sides of the SI, we do not see a sudden dip in the overall particle intensity near the SI, something that is often observed in energetic particle events related to CIRs \citep[see e.g.][]{strauss16, dwyer97}. 
 However, as described in Section~\ref{sec:observers}, near the SI we do see a decrease in intensity of the high-energy channels, since the
magnetic field lines immediately on either side of the SI originate from the transition region itself.
 These IMF lines therefore do not intersect the compression or shock waves bounding the CIR, where the particles may accelerate or reflect.
 We also note that in our simulations, the reason why the SI acts as a diffusion barrier is because of the increase in magnetic field strength in the shocked slow solar wind.
  However, \cite{strauss16} argue that the reduction in cross-field diffusion near the SI is due to strong damping of magnetic field fluctuations perpendicular to the SI, leading to an anisotropic cross-field diffusion. 
  
Finally, we note that at 1 AU the ratio of the perpendicular mean free path to the parallel   mean free path in the CIR is, in our simulations, at most $\lambda_\perp/\lambda_\parallel = 3.81\times10^{-4}$. This is small compared to the values obtained by \cite{dwyer97}, who find ratios of the order of unity for three CIRs  using data from the Wind spacecraft. 
 However, in our results  below we show that the magnetic field topology inside a CIR may  increase the effect of the cross-field diffusion, without necessitating increased amounts of turbulence or large values of $\lambda_\perp/\lambda_\parallel $.

Figure~\ref{fig:all_particles} shows the total particle density drawn in grey shades  on top of the radial velocity profile of the background solar wind in the solar equatorial plane, 16~hours after the particle injection.
 The left panel of  Fig.~\ref{fig:all_particles} corresponds to the simulation assuming $\kappa_{\perp} = 0$, while the right panel corresponds to the simulation with $\kappa_{\perp} \ne 0$. 
In contrast to the cases discussed in Section~\ref{sec:parker_seps} with a Parker solar wind configuration,  the difference between the simulations is significant. 
 At $1$ AU, the longitudinal extent of the particle zone for the case with cross-field diffusion is more than twice that of the case without cross-field diffusion, although this extended area shows a particle density two orders of magnitude smaller than in the main streaming zone. We also performed a simulation where we injected particles only in the slow solar wind of the EUHFORIA simulation and obtained similar results as for as for the simulations shown in Section~\ref{sec:parker_seps}, that is, the cross-field diffusion had only a weak effect.
The reason why the cross-field diffusion becomes more effective in the CIR  is because the CIR contains compressed IMF lines that are widely separated in the unperturbed  solar wind (see Fig.~\ref{fig:all_particles}). An example of such magnetic field lines is the pair of orange lines in Fig.~\ref{fig:all_particles}.  Even a very small cross-field motion in the CIR can transport a particle across these magnetic field lines.
 This results in a significant angular spread of the particle density for $r\lesssim 2$~AU, when particles return from the CIR to the unperturbed fast or slow solar wind. 
 Such a reversal in the propagation direction of particles travelling inside the CIR is facilitated due to the IMF lines converging towards the SI, hence acting as a magnetic mirror. 
Apart from the SI, the forward and reverse shock waves may also mirror the particles.  
As a consequence of the existence of these three magnetic mirrors, there is a significant amount of sunward propagating particles and particle densities remain high at small radial distances for a prolonged amount of time, as shown in Section~\ref{sec:observers}.
 
\subsection{Particle acceleration}\label{sec:acc}
  \begin{figure*}
        \centering
        \includegraphics[width=0.9\textwidth]{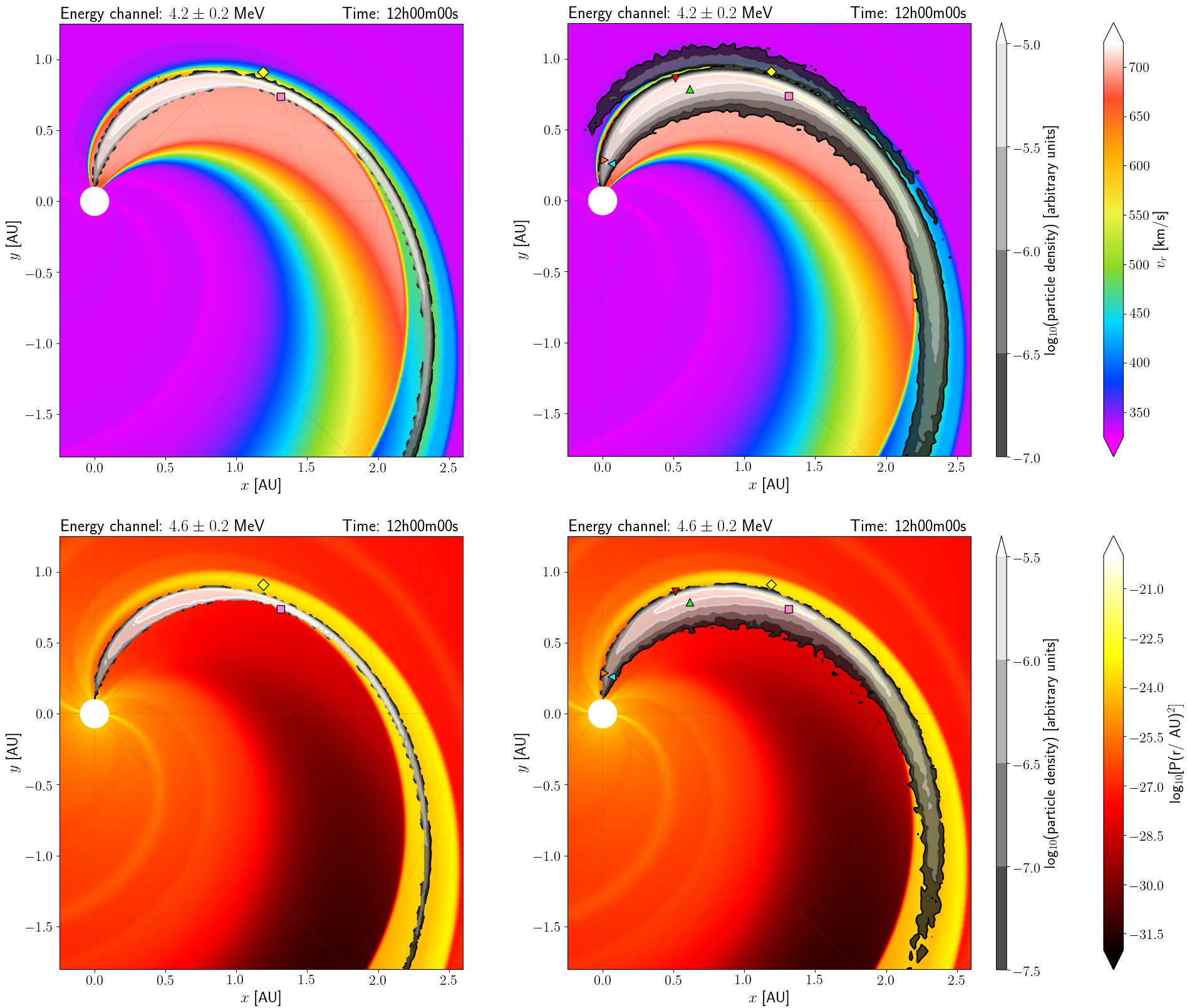}
    \caption{Top row shows particle intensities of energy channel $4.2\pm 0.2$ MeV drawn in grey shades on top of the radial velocity profile of the solar wind. The bottom row shows particle intensities of energy channel $4.6\pm 0.2$ MeV drawn in grey shades on top of scaled pressure profile of the solar wind. The right and left  columns correspond, respectively, to the simulations with and without perpendicular diffusion.  The entire temporal evolution is available as an online movie. }
     \label{fig:euh_acc_channels}
\end{figure*}
The FTE contains the necessary physics for modelling the acceleration of particles travelling in  converging or accelerated flows for example.
 The different mechanisms that alter the energy of the  particles can be more easily understood by rewriting Eq.~\eqref{eq:fte_p} in the following form \citep{leRoux12}:
\begin{equation}\label{eq:dpdtLeRoux}
\frac{1}{p}\td{p}{t} = -\frac{1}{3}\nabla\cdot\vec{V_{sw}} + \frac{1}{2}(1-3\mu^2)\vec{bb}:\bm{\sigma} - \frac{\mu}{\varv} \vec{b}\cdot\td{\vec{V_{\rm{sw}}}}{t},
\end{equation}  
where $\bm{\sigma}$ denotes the shear tensor given by
\begin{equation}
\sigma_{ij} = \frac{1}{2}\left(\pd{V_{\rm{sw},i}}{x_j} + \pd{V_{\rm{sw},j}}{x_i} - \frac{2}{3}\pd{V_{\rm{sw},i}}{x_j} \delta_{ij}\right).
\end{equation}
The first term on the right hand side of Eq.~\eqref{eq:dpdtLeRoux} illustrates the effect of converging or diverging flows on the energy of the  particles.  
A particularly important example of a converging flow is found at a shock wave, where the plasma flow has a negative divergence. 
 Each time the particles cross the shock they will therefore accelerate, a mechanism known as diffusive shock acceleration or first-order Fermi acceleration.
 As discussed in Section~\ref{sec:cir}, the EUHFORIA simulation contains two separate shock waves bounding the CIR which, as shown below, accelerate particles. 
Such diffusive particle acceleration at the forward and reverse CIR shock waves is considered as a likely mechanism to explain the particle intensity peaks often measured at CIRs \citep[see e.g.][and references therein]{richardson04}.
As already noted before, the finite resolution of the MHD simulation will unavoidably smear out the shocks over a spatial region  larger than the width of real CIR shocks, making the terminology ``shock acceleration'' not strictly applicable to our results.   
However, as long as the particle mean free path across the shock is much larger than the width of the shock, the particles will gain energy due to their motion and scattering in rapidly converging flows, similar to what happens during first-order Fermi shock acceleration \cite[see also ][for a discussion]{Giacalone02}. 
For the same reason, particles will gain energy when crossing the large amplitude compression waves bounding the transition region between the slow and fast solar wind at small radial distances, before the CIR  and its bounding shocks have formed.  
These  compression waves are indeed also characterised by a negative flow divergence, and hence particles will gain energy when crossing them due to the interaction with converging scattering centres.
This mechanism was proposed by \cite{Giacalone02} as the main acceleration mechanism in CIRs, producing energetic particle populations at small radial distances, before the forward and reverse shocks are formed.

In our simulation,  IMF lines to which the majority of the particles are injected are initially diverging while residing in the unperturbed slow or fast solar wind, and converging once entering the transition zone or CIR at larger radial distances.
 Hence,  particles injected on those IMF lines will initially lose energy due to adiabatic deceleration, and will later gain energy due to adiabatic acceleration.
 In addition,  there are also particles injected in the transition zone itself that therefore follow IMF lines which never cross any compression or shock wave and that are immediately  adjacent to the SI in the CIR. 
Due to the lack of any compression or shock wave crossings, these particles will not experience any substantial acceleration. However, inside the CIR the flows are slowly converging towards the SI, which may result in  weak adiabatic acceleration.  

The other two mechanisms contained in Equation~\eqref{eq:fte_p} that alter the momentum of the particle are the flow shearing and flow acceleration, respectively represented in the  last two terms on the right hand side of Eq.~\eqref{eq:dpdtLeRoux}. 
These terms are non-zero at the shock waves, and it is the cosine of the pitch angle which determines whether the particle gains or loses energy upon crossing the shock.
For example, the acceleration (deceleration) of the solar wind across the forward (reverse) shock will increase (decrease) the energy of a particle when it travels from the downstream to the upstream region, and vice-versa.  
We note that there is also  flow shear in the CIR, especially at the streaming interface, where the shocked decelerated fast solar wind meets the shocked accelerated slow solar wind.  

Figure~\ref{fig:euh_acc_channels} shows, at $t = 12$~hours, the  particle density of protons accelerated above their initial energy, registered in two energy channels: $4.2\pm 0.2$ MeV (top row) and $4.6\pm 0.2$ MeV (bottom row). The left (right) column shows the results for the simulation without (with) cross-field diffusion.  
The background in the top and bottom row depict, respectively, the radial velocity and the scaled pressure, $r^2P$, of the  solar wind.
The four panels of Figure~\ref{fig:euh_acc_channels} clearly show the existence of an accelerated particle population centred on the reverse shock wave.  
As seen in the temporal evolution of this particle population  (available as an online movie), the $4.2\pm 0.2$ MeV channel starts being populated $2$ hours and $40$ minutes after the injection, and at a radial distance of ${\sim}1.2$ AU, while the $4.6\pm 0.2$ MeV channel starts getting populated about 1 hour and 40 minutes later, at a slightly larger radial distance (${\sim}1.35$ AU).
 
By calculating the length of the field line on which the first accelerated particles appear, a particle of 4 MeV would need ${\sim}2.2$~hours  to reach the acceleration zone if it were to travel scatter-free.
 When travelling towards the reverse shock, the particles will however lose momentum due to adiabatic deceleration, and hence will arrive at the shock wave with an energy lower than 4 MeV.
As shown in Section~\ref{sec:parker_seps}, this energy loss can be significant, especially since the particles are travelling in a fast solar wind during their journey towards the reverse shock wave.
Therefore, the particles may need more than one shock crossing to reach energies above  4 MeV. 
 Multiple shock wave crossings are facilitated due to the focusing/mirroring effect at both sides of the shock.
 When travelling in the sunward direction, particles will eventually be mirrored and focused in the direction of the shock as a consequence of the converging magnetic field lines near the sun. 
After crossing the shock wave, particles will propagate in the CIR where the magnetic field lines are converging towards the SI, and hence particles are likely to eventually get mirrored and focussed back towards the shock. 
  Apart from particle mirroring due to the mean magnetic field,  scattering of the particles due to turbulence will also facilitate multiple shock crossings.  

A notable difference between the simulations with and without cross-field diffusion is the existence of a second population of accelerated particles for the former case. 
Centred on the forward shock of the CIR, this population is only visible in the $4.2\pm 0.2$ MeV energy channel (see the top right panel of Fig.~\ref{fig:euh_acc_channels}) and from ${\sim} 6$ hours after  the particle injection, at a radial distance of ${\sim}1.6$ AU.
The IMF lines originating from the injection zone do not traverse the forward shock at these large radial distances, and hence these particles are there as a result of cross-field motions inside the shocked slow solar wind. We further note that  both populations of accelerated particles are separated by the SI.
The existence of this additional population of accelerated particles  illustrates that a weak cross-field diffusion can have noticeable effects on energetic particle populations when the particles are travelling in a non-nominal solar wind configuration.

\subsection{Proton fluxes measured by virtual observers}\label{sec:observers}

\begin{figure*}
        \centering
        \begin{tabular}{cc}
        \includegraphics[width=0.4\textwidth]{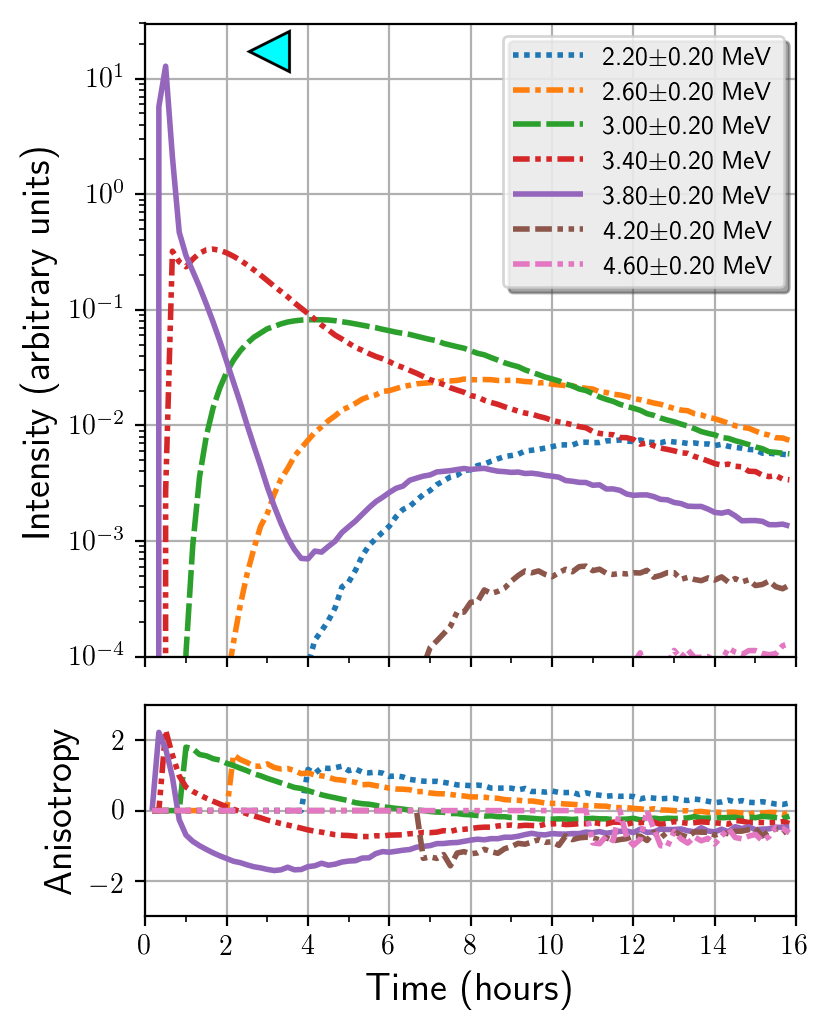}&
        \includegraphics[width=0.4\textwidth]{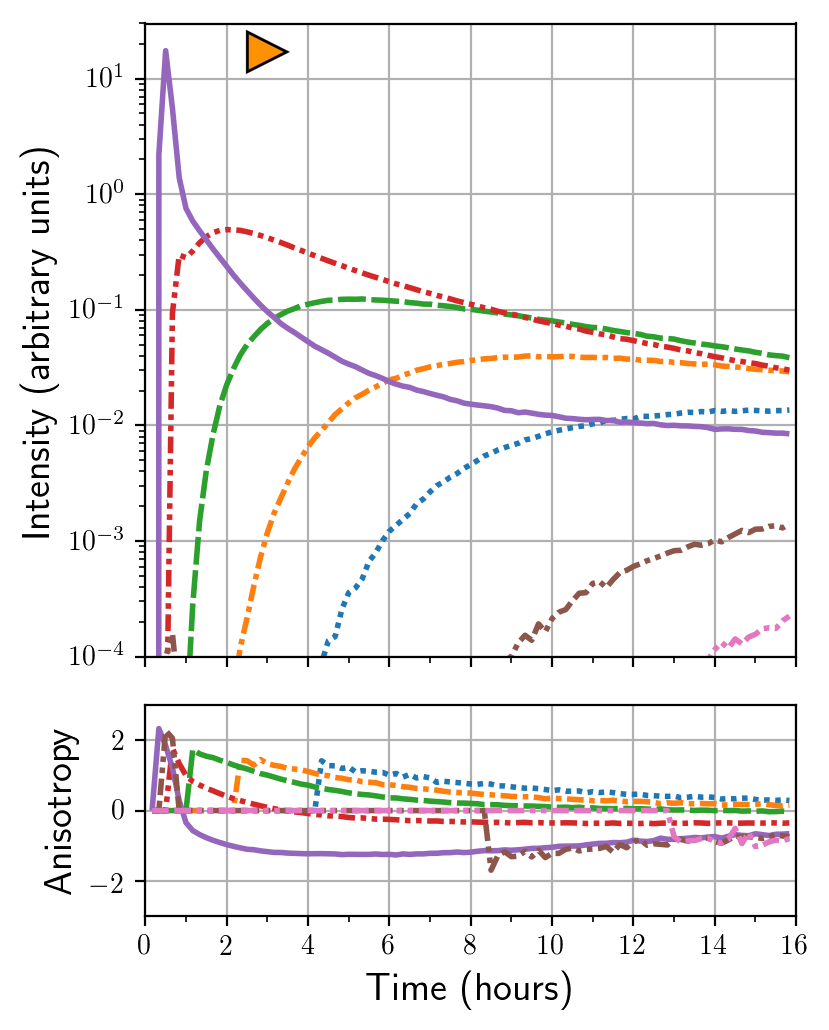}\\
    \end{tabular}
    \caption{Proton intensities (top panels) and anisotropies (bottom panels) measured in the solar equatorial plane at a heliocentric radial distance of 0.28 AU and at a longitudes of $76.5^\circ$ (left panel) and $87.5^\circ$  (right panel).}
     \label{fig:tp0.28AU}
\end{figure*}
We now analyse the proton intensities and anisotropies measured by a fleet of virtual observers located in the solar equatorial plane, covering the entire zone where energetic particles travel during the first 16 hours after their injection. 
We have studied 47 stationary observers positioned at radial distances of 0.28~AU, 1~AU, 1.5~AU and 2~AU from the Sun. In order to illustrate the richness and variety of different time profiles encountered, even when looking at observers located relatively close to each other, we have selected six observers to be discussed.
 These observers are indicated by the various markers in Fig.~\ref{fig:all_particles}. 
All the intensities are normalised to the maximum intensity measured at 1 AU in the $3.8\pm 0.2$ energy channel by the observer located at $56.11^\circ $ in longitude (not shown here).
In this section, we mainly focus on the simulations done with cross-field diffusion, except for the observers located at 1.5 AU, for which we discuss also the case $\kappa_\perp = 0$

Figure~\ref{fig:tp0.28AU} shows  2.0\,--\,4.8~MeV proton intensity-time profiles (top panels) and anisotropy time profiles (bottom panels) gathered by two observers located at a radial distance of $0.28$ AU, which corresponds to the closest planned perihelion for Solar Orbiter. 
The left panel of Fig.~\ref{fig:tp0.28AU} shows the results for the observer located at a longitude of $76.5^\circ$ represented by the cyan left-facing triangle in Fig.~\ref{fig:all_particles}.
This observer is positioned in the fast solar wind during the entire simulation. 
Initially, at $t = 0$~hours, the observer is magnetically connected to the reverse shock at $r\sim1.2$ AU, yet we note that this connection point moves towards larger radial distances due to the corotation of the reverse shock with the Sun.  
The early sharp intensity peak of the $3.8\pm 0.2$~MeV energy channel (solid purple curve) reflects the mono-energetic delta time injection of the 4 MeV protons. 
However, the $3.4\pm 0.2$~MeV energy channel (dot-dashed red curve) and even the lower-energy channels are quickly populated due to efficient adiabatic deceleration, since particles travel in a fast wind stream.  

A particularly interesting feature is that the intensity of the $3.8\pm 0.2$ energy channel switches from rapidly decreasing to gradually increasing after ${\sim}4$ hours. 
This increase is due to particles that got mirrored at the reverse shock or in converging IMF lines inside the CIR. 
The reason why these particles have not yet adiabatically decelerated to lower energy channels is because they have gained energy at the reverse shock, counteracting the energy losses they suffer when travelling in the fast solar wind.

The right panel of Fig.~\ref{fig:tp0.28AU} shows the intensity and anisotropy profiles for an observer located at a longitude of $87.5^\circ$. Like in the previous instance, this observer is positioned in the fast solar wind, but this time close to the transition region towards the slow solar wind (see the right-facing orange triangle in Fig.~\ref{fig:all_particles}). 
The main difference with respect to the former observer is that the  intensity profile of the $3.8\pm 0.2$~MeV protons does not show this double peaked structure, but instead  it monotonically decreases after onset. 
This is because the IMF lines connecting the observer move already at small radial distances ( $< 1$~AU) into the transition zone between the fast and slow solar wind, and hence the magnetic mirrors are much closer to the observer.
 At these distances, the forward and reverse shock waves are not yet formed and hence there is no significant acceleration of particles. 
 Despite the absence of a real shock wave, the particles may adiabatically gain energy upon crossing the compression wave between the fast solar wind and the transition zone, keeping the $3.8\pm 0.2$~MeV energy channel populated.  
We conclude therefore that the combination of acceleration and mirroring  at the compression/shock waves explains  why, for both observers at 0.28 AU, the intensities in all energy channels remain high throughout the simulation, which is in sharp contrast with the intensity profiles shown in the lower-left panel of Fig.~\ref{fig:DecParker} for an observer in a Parker spiral at a similar radial distance.
\begin{figure*}
        \centering
        \begin{tabular}{cc}
        \includegraphics[width=0.4\textwidth]{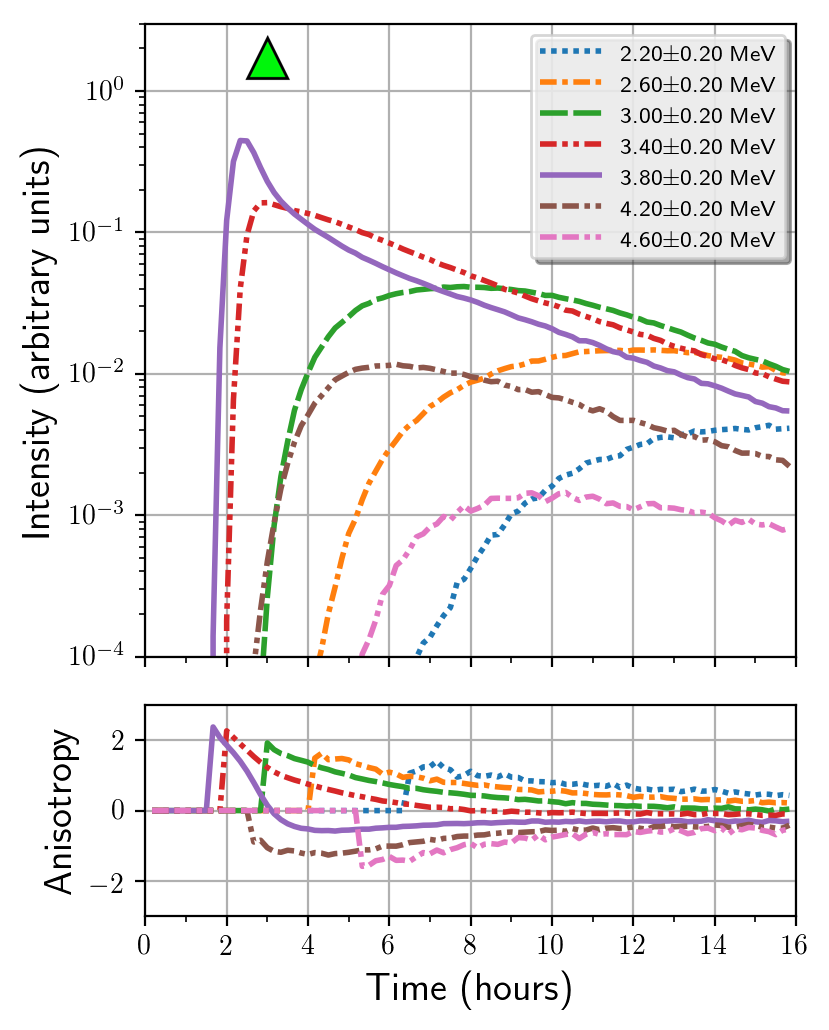}&
        \includegraphics[width=0.4\textwidth]{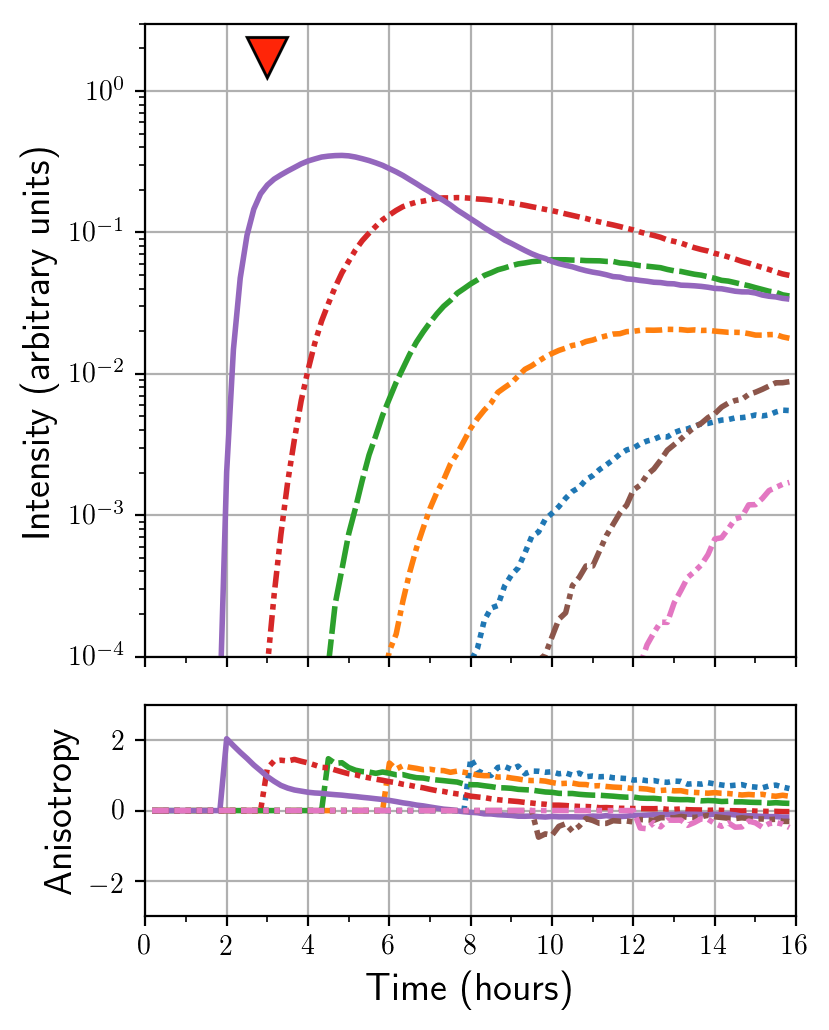}\\
    \end{tabular}
    \caption{Proton intensities (top panels) and anisotropies (bottom panels) measured in the solar equatorial plane at a heliocentric radial distance of 1 AU and at a longitudes of $52^\circ$ (left panel) and $59.3^\circ$  (right panel).}
     \label{fig:tp1AU}
\end{figure*}
Next we consider two observers located at a radial distance of $1$~AU. The first of these observers is located at a longitude of $52^\circ$ (green upright triangle in the right panel of Fig.~\ref{fig:all_particles}).
The proton intensity and anisotropy profiles for this observer are shown in the left panel of Fig.~\ref{fig:tp1AU}. 
As for the observers at 0.28~AU, particles propagate towards this observer in the fast solar wind and hence low-energy channels close to 4~MeV are populated rapidly due to adiabatic deceleration. 
During the entire simulation, the IMF lines passing through the stationary observer cross the reverse shock at a radial distance of around 1.3 AU. 
The observer is thus constantly closely connected to the reverse shock at the location where the first population of accelerated particles is generated (see Section~\ref{sec:acc}). 
As a consequence, this observer measures relatively high particle intensities in the $4.2\pm 0.2$~MeV (brown curve) and $4.6\pm 0.2$~MeV (pink curve) energy channels, which are populated by those shock-accelerated particles. 
Since the observer is connected to the reverse shock at a radial distance larger than 1 AU, and the magnetic field is pointing away from the sun, the anisotropies of the accelerated particles are all negative. 
We note that after ${\sim}3$~hours, the anisotropy for the $3.8\pm 0.2$~MeV channel (purple curve) is negative, suggesting that this channel is also mainly populated  by particles that interacted with the shock.   
This explains the slow decrease of the $3.8\pm 0.2$~MeV proton intensities after ${\sim}3$ hours which is in sharp contrast to the intensity profiles discussed in Section~\ref{sec:parker_seps} for particles travelling in a Parker solar wind (see e.g. the lower-right panel of Fig.~\ref{fig:DecParker} for comparison). 

The second observer at $1$~AU is located at a longitude of $59.3^\circ$ (see the downward-facing red triangle in the right panel of Fig.~\ref{fig:all_particles}).
Initially, this observer is positioned on the compression wave between the slow solar wind and the transition region.
Later on, the transition region rotates past the observer until it crosses the boundary between the transition region and the fast solar wind
near the end of the simulation. 
The intensity and anisotropy profiles for this observer are shown in the right panel of Fig.~\ref{fig:tp1AU}. 
The most remarkable feature is the distinct shape of the $3.8\pm 0.2$ MeV energy channel, and the relatively big time gap between the arrival of particles of energy channels $3.8\pm 0.2$ MeV and  $3.4\pm 0.2$ MeV, as compared to the first observer at 1 AU (see the left panel of Fig.~\ref{fig:tp1AU}).
 These features can be explained as a consequence of the interplay between adiabatic acceleration and deceleration as follows. 
The first particles reaching the observer at ${\sim} 2$~hours are following IMF lines that start in the slow solar wind, and hence they are adiabatically decelerated during the first part of their journey.
 However, as explained in Section~\ref{sec:parker_seps},  this adiabatic deceleration will only be limited due to the slow wind speed (${\sim} 330$  km~s$^{-1}$). 
Before reaching the observer, the particles leave the slow solar wind and cross the compression wave bounding the transition zone, where they may accelerate to some extent.  
 The net result is that more particles remain in their initial $3.8\pm 0.2$~MeV energy channel.
  After ${\sim} 5$ hours, the intensity of the $3.8\pm 0.2$~MeV energy channel starts decreasing more rapidly until ${\sim} 10$ hours. 
During that decrease, the observer is connected to IMF lines that are (almost) entirely inside the transition region at small radial distances and at larger radial distances they lie adjacent to the SI.
 Hence, as explained in Section~\ref{sec:acc}, due to the lack of converging flows, the particles following these field lines will mostly adiabatically decelerate. In addition, these particles do not experience the magnetic mirroring effect of the compression or shock waves. 
 Later on, after ${\sim} 10$ hours, the intensity of the $3.8\pm 0.2$~MeV energy channel starts decreasing more gradually, since the observer becomes connected to IMF lines that cross the compression wave  between the fast solar wind and the transition zone, where the particles may accelerate.   
Near ${\sim} 12$~hours, the observer enters the fast solar wind and  establishes connection with IMF lines crossing the reverse shock at radial distances larger than 1~AU. 
 We note that the observer already receives particles with energies above 4~MeV before 12~hours. However these particles are there solely due to cross-field diffusion since they are not observed in the simulations with zero cross-field diffusion (not shown here). 
 \begin{figure*}
        \centering
        \includegraphics[width=0.8\textwidth]{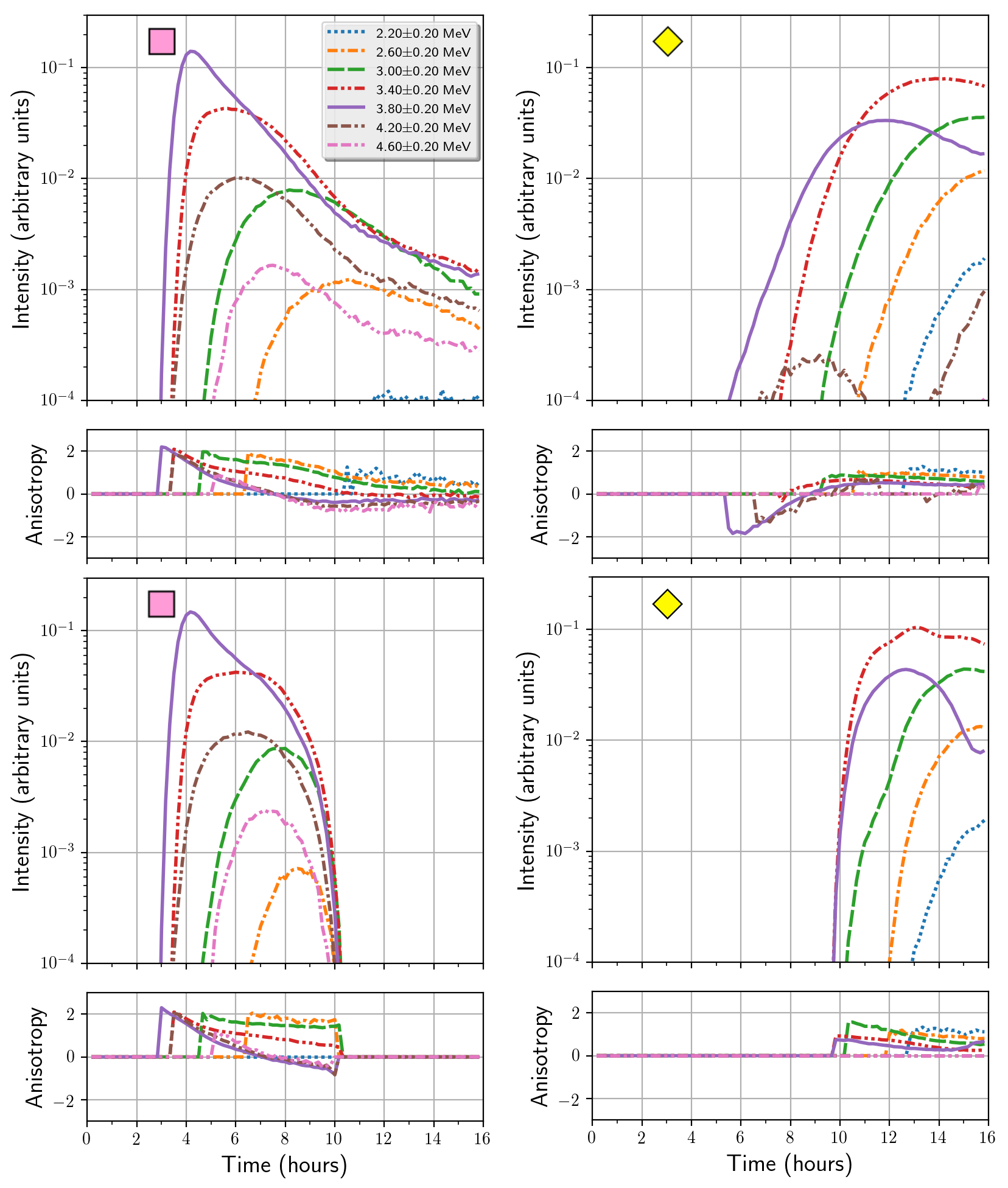}
    \caption{Intensities and anisotropies measured in the solar equatorial plane at a  heliocentric radial distance of 1.5 AU and at longitudes of $29.8^\circ$ (left column) and $37.54^\circ$  (right column). The top and bottom rows correspond, respectively, to the simulations with and without perpendicular diffusion.}
     \label{fig:tp1.5AU}
\end{figure*}
\begin{figure*}
        \centering
        \begin{tabular}{ccc}
        \includegraphics[width=0.31\textwidth]{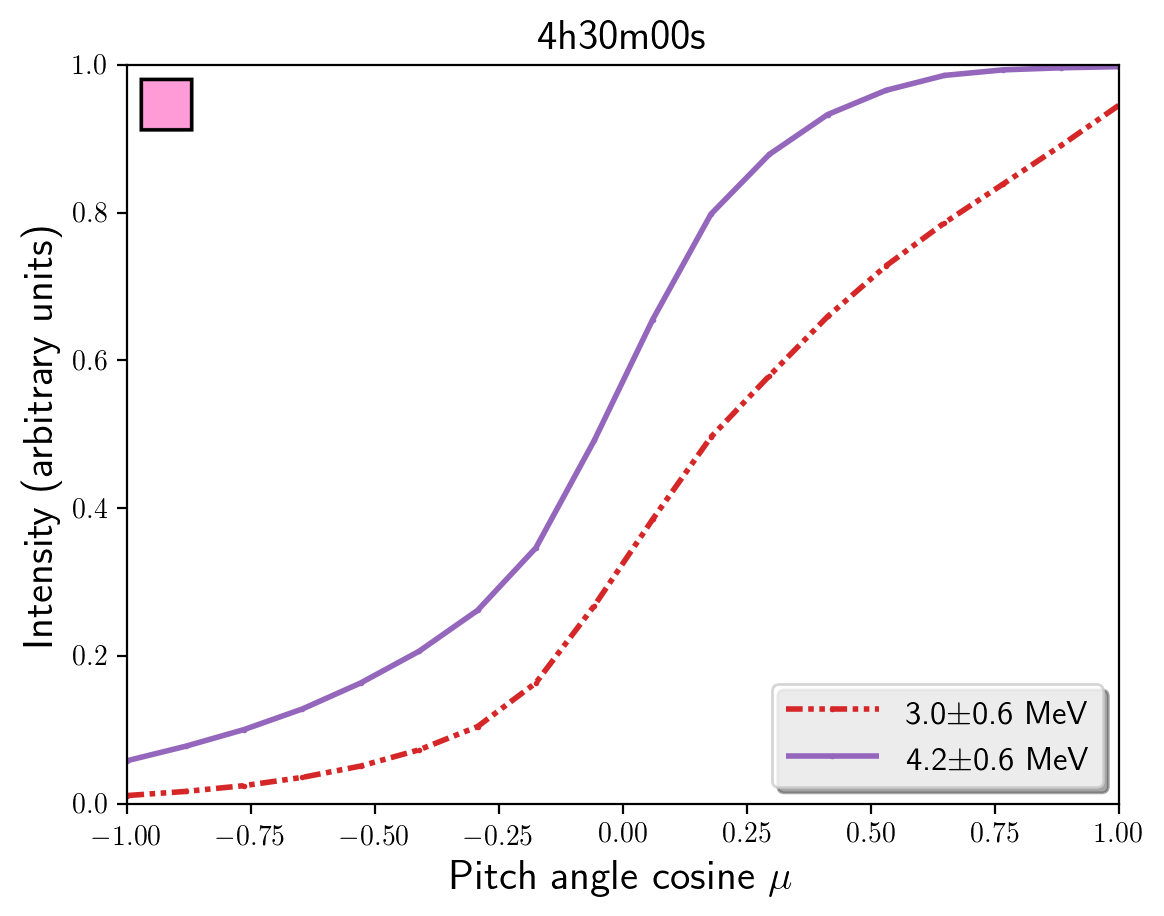}&
        \includegraphics[width=0.31\textwidth]{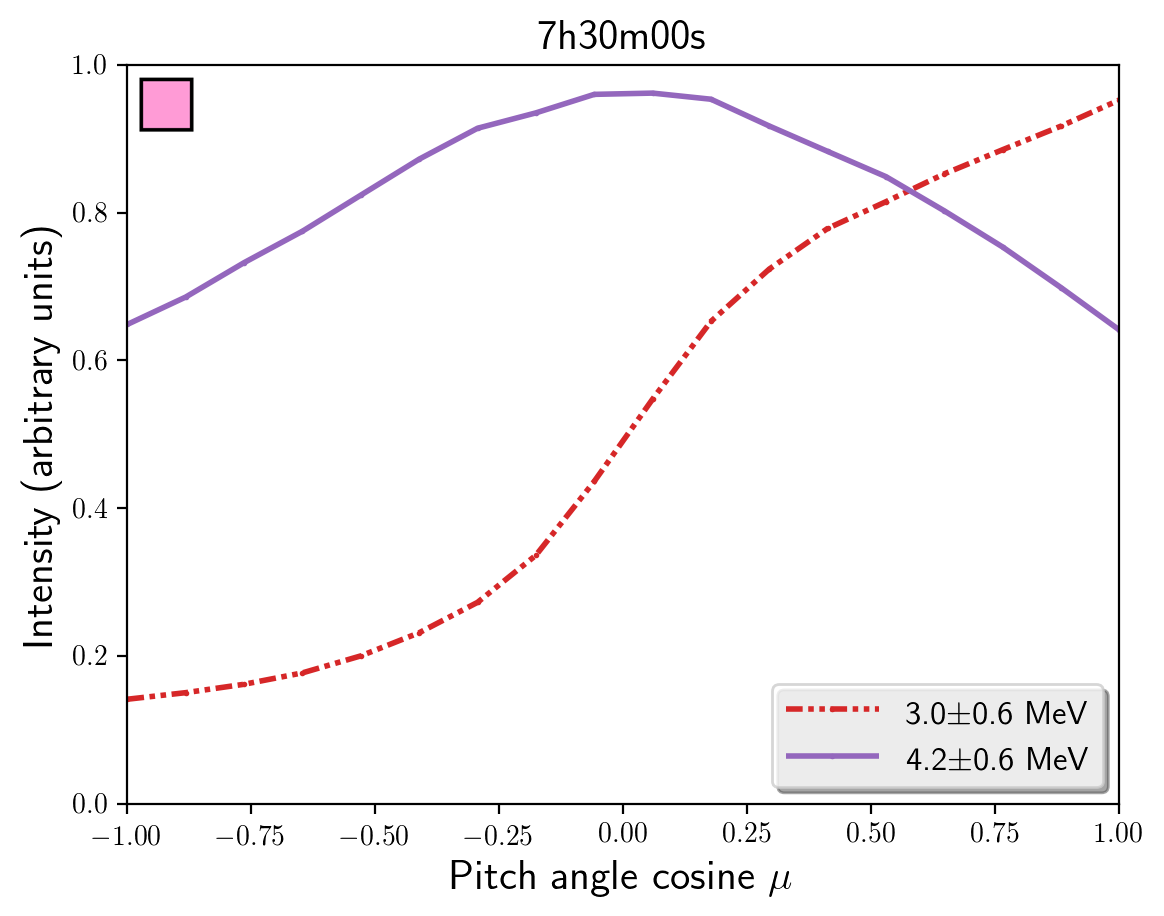}&
        \includegraphics[width=0.31\textwidth]{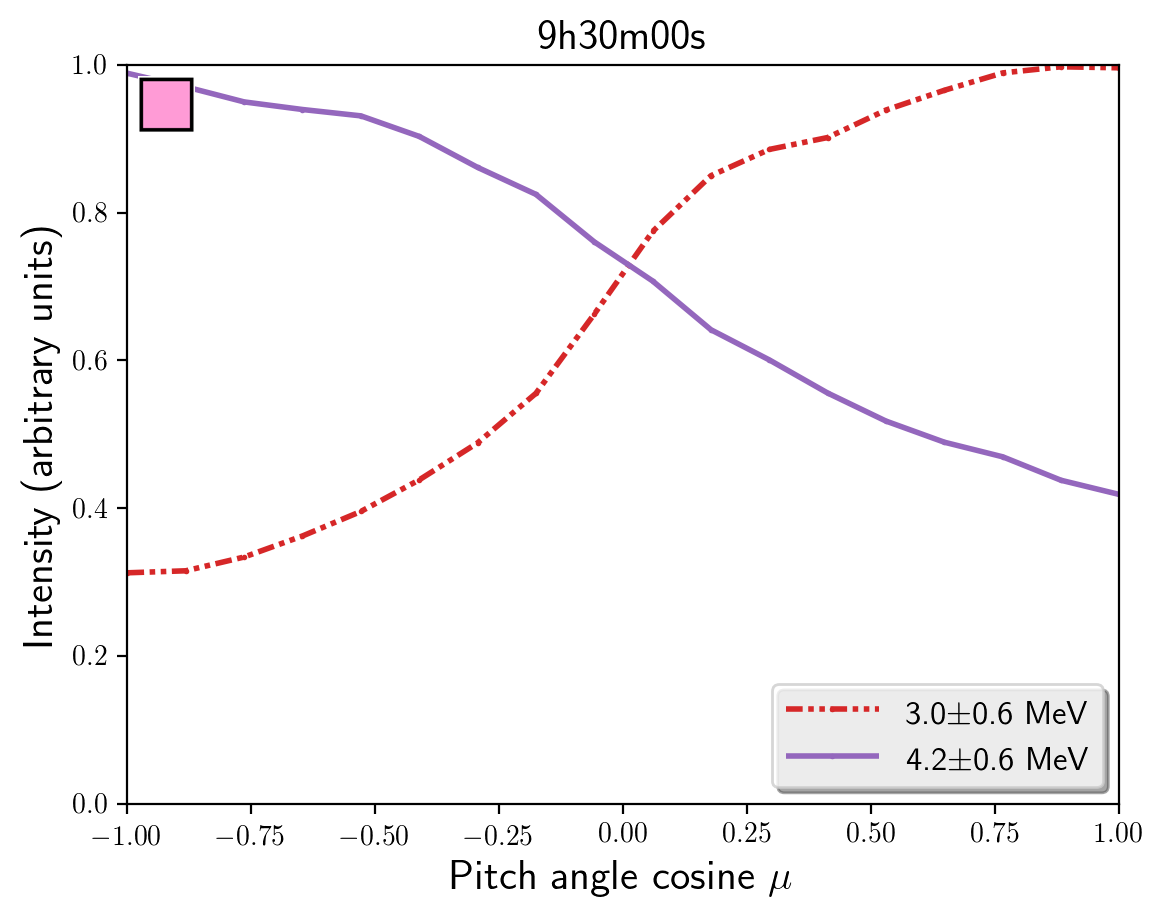}\\
    \end{tabular}
    \caption{Pitch-angle distribution functions at three different time instances for an observer located at  1.5 AU and at a longitude of $29.8^\circ$. See top-left panel of Fig.~\ref{fig:tp1.5AU} for the intensities and anisotropies of this observer. }
     \label{fig:pad1.5AU}
\end{figure*}

For the two observers located at $1.5$ AU, we discuss the resulting profiles for both the simulations with and without cross-field diffusion. 
The first observer we consider (pink square in Fig.~\ref{fig:all_particles}) is located at a longitude of $29.8^\circ$.  
The top (bottom) graphs in the left column of Fig.~\ref{fig:tp1.5AU} show the particles' intensity and anisotropy profiles seen by this observer for the case with (without) cross-field diffusion.
This observer is particularly interesting because it crosses the particle acceleration site at the reverse shock, after ${\sim}$7 hours and 30 minutes. 
This is indicated by the anisotropies of the energy channels containing accelerated particles which switch sign around that time. This becomes clear after inspecting the pitch-angle distributions (PADs). Figure~\ref{fig:pad1.5AU} shows, for the case with cross-field diffusion, the proton PADs at three time instances for two different energy channels, $4.2\pm 0.6$~MeV (purple curves) and $3.0\pm 0.6$~MeV (red curves). 
The left panel corresponds to 4 hours and 30 mins after the injection, when the shock is between the Sun and the observer, whereas the right panel shows the PADs at 9 hours and 30~minutes, when the observer is located between the Sun and the shock. 
The middle panel shows the intermediate situation, that is, when the observer is located on the shock. The $4.2\pm 0.6$~MeV channel is mainly populated with shock-accelerated particles and hence, the corresponding PAD (purple curves) turns over during the shock passage, i.e. the PAD evolves from an increasing to a decreasing function of $\mu$.
 At the shock passage, the PAD becomes more horizontal, reflecting the less anisotropic conditions there. In contrast, the $3.0\pm 0.6$~MeV channel is mostly populated by particles adiabatically decelerated at smaller radial distances. Therefore, the PAD of this energy channel remains an increasing function of $\mu$ during and after the observer crossing the shock.

We also note that in Fig.~\ref{fig:tp1.5AU},  the $2.2\pm 0.2$~MeV channel is almost completely depleted of any particles for this observer. 
This is in contrast to what is observed at 0.28~AU and at 1~AU, where this energy channel shows significant intensities.  
This difference can be attributed to the combined effect of shock acceleration and adiabatic compression in the CIR occurring at radial distances larger than 1 AU. 

The drop in intensities after ${\sim}10$~hours is because the observer moves out of the particle streaming zone due to the corotation effect, as  is clearly seen in the case without cross-field diffusion (bottom left panel of Fig.~\ref{fig:tp1.5AU}).
At the same time, the top panel of  Fig.~\ref{fig:tp1.5AU} shows that the cross-field diffusion has a stronger effect compared to the cases discussed in Section~\ref{sec:parker_seps}.  

The second observer at $1.5$ AU is located at a longitude of $37.54^\circ$ (yellow diamond in Fig.~\ref{fig:all_particles}).  
This observer is initially positioned in the slow solar wind outside the particles' streaming zone, and only enters this zone after $> 6 $~hours due to the corotation of the particles with the IMF, as shown by the intensity and anisotropy time profiles in the right column of Fig.~\ref{fig:tp1.5AU}.
Comparing the cases with (top panel) and without (bottom panel) cross-field diffusion, the onset of the particle event occurs about 4 hours earlier in the former case.
For this case, the first observed particles are all high-energy particles (channel $3.8\pm 0.2$~MeV) showing a negative anisotropy. 
This agrees with the finding above that the cross-field diffusion is more effective at large radial distances in the CIR, such that high-energy particles are initially affected more strongly because they travel faster to those  distances.
This also explains the time-gap between the arrival of particles populating the $3.8\pm 0.2$ MeV and  $3.4\pm 0.2$ MeV channels.  The negative anisotropy is again an indication of the mirroring effects of the CIR.

Also of interest is the first small bump of $4.2\pm 0.2$~MeV protons seen only in the simulation with cross-field diffusion, around 9 hours. This bump is due to particles accelerated by the forward shock, and corresponds to the second accelerated particle population discussed in Section~\ref{sec:acc} that reached those IMF lines due to cross-field motions. At the end of the particle event, this observer detects a second increase of intensity in the $4.2\pm 0.2$~MeV channel  for the case with cross-field diffusion, since the observer approaches the reverse shock wave. 

For the simulation assuming $\kappa_\perp=0$ (see the bottom right panel of Fig.~\ref{fig:tp1.5AU}), after a prompt onset, the $3.8\pm 0.2$ MeV proton intensity-time profile shows a rounded plateau-shape similar to that obtained for the observer at 1 AU depicted in the right panel of Fig.~\ref{fig:tp1AU}. 
This shape can again be  attributed  to the non-trivial interplay between acceleration/deceleration processes occurring in the slow solar wind, the transition zone and the forward shock. 
The sharp decrease  between 13 and 15  hours is again due to the passage of the SI and its adjacent magnetic field lines, that is, the field lines that do not cross any shock/compression wave where the particles can accelerate and mirror.

\section{Summary and Conclusions}\label{sec:summary}
In this article, we present and test a new particle transport code that obtains solutions of the focused transport equation by means of a Monte-Carlo simulation. 
The focused transport equation that we solve is extended from the standard formulation by including for the first time both the effects of the  magnetic gradient/curvature drifts and cross-field diffusion.
 The new code is a fully parallelized 3D time-dependent particle transport code, able to propagate particles in complex solar wind configurations generated by 3D MHD models like EUHFORIA.  

In Section~\ref{sec:parker_seps} we tested the code by propagating particles in a nominal IMF for slow and fast solar wind configurations. 
We started with presenting the results of simulations  both with and without cross-field diffusion, yet neglecting particle convection and adiabatic energy losses in the solar wind. 
These simulations illustrated how cross-field diffusion can make sharp cut-offs in particle intensities more gradual, a result previously shown by \cite{droge10}. 
Subsequently we included all terms of the FTE, and illustrated how particles are considerably more adiabatically decelerated in the fast solar wind than in the slow solar wind. 
For a fixed source of particles near the Sun, \cite{ruffolo95} and \cite{kocharov98} quantified the decay rate of proton intensities for 1 AU observers due to adiabatic deceleration. We have shown instead the substantial energy loss of particles in the fast solar wind by depicting how the 4~MeV injected protons populated lower-energy channels in the case of observers located at two different radial distances. 
Comparing, for those observers, the most populated energy channels at every time instance reveals differences that reflect the dependence of magnetic focusing on the velocity of a particle and on the radially decreasing magnetic field strength. 

In the second part of the article we propagated particles in a solar wind generated by EUHFORIA.
In particular we modelled a slow solar wind configuration with an embedded fast solar wind stream. 
The  substantial difference between the slow and fast solar wind speeds resulted in the formation of a CIR bounded by a forward and a reverse shock at relatively small radial distances (${\sim}1.5$ AU onwards). 
Hence, unlike previous simulations of impulsive SEP events in CIRs \citep{Giacalone02,kocharov08a}, we use a 3D MHD simulation that includes both the forward and reverse shock waves in the same set-up, which allows a more complete description of the particles' intensity-time profiles obtained by virtual observers placed at different locations in the ecliptic plane.
We considered an impulsive injection of 4 MeV protons at the inner boundary of EUHFORIA, uniformly spread over a region covering a small part of the slow solar wind and a substantial part of the fast solar wind stream. 
The particles were propagated both with and without cross-field diffusion. 
Despite using a small perpendicular mean free path, the differences between both cases were substantial. 
When cross-field diffusion was switched on,  particles spread over a much larger region in the heliosphere, more than doubling its longitudinal extent in the solar equatorial plane. 
In contrast, this feature was not observed in  the simulations using a simple Parker solar wind, despite using a similar cross-field diffusion. 
Therefore, the increase of efficiency of the cross-field diffusion in the simulation containing the embedded fast solar wind stream can be attributed to the more complex magnetic field configuration  found in the CIR. 
In particular, inside the CIR,  magnetic field lines are converging such that  small cross-field motions can transport particles to magnetic field lines that are widely separated in the unperturbed solar wind. This increase of efficiency of cross-field diffusion at the boundary between a slow and fast solar wind  stream  could potentially help explain some of the actual measured particle events that show a large angular spread in the heliosphere. 

We note that we are using a rather simplified model for the cross-field diffusion. 
However since we use a very weak cross-field diffusion, which is in addition minimal at the CIR shocks and at the SI due to its inverse scaling with the magnetic field strength, it is likely that a more realistic  treatment of the cross-field motions will influence the particle densities in the heliosphere in a similar or even stronger way.

Our simulations also show the formation of an accelerated particle population centred on the reverse shock of the CIR.
 In particular, the acceleration site of the particles is mainly situated at a radial distance of ${\sim}1.5$ AU, yet this location merely reflects the magnetic connection between the reverse shock and the particles injection region. 
There are also field lines originating from the  particle injection region that cross the boundary between the slow or fast solar wind and the transition zone at small radial distances, i.e. before the compression waves have steepened into shock waves.
The particles following these field lines already show signatures of strongly reduced adiabatic deceleration and even adiabatic acceleration.
This is exemplified
by the plateau shape in the $3.8\pm 0.2$ MeV energy channel during the first hours of the particle event (see, e.g. the right panel of Fig.~\ref{fig:tp1AU}).
A second population of accelerated particles, centred on the forward shock near ${\sim}1.6$ AU, appeared in the case of the simulation with cross-field diffusion. 
This population of accelerated particles was not present in the case without cross-field diffusion, since no magnetic field lines connected the particle injection region with the forward shock at those larger radial distances. 
Remark that these forward shock accelerated particles were injected in the slow solar wind, since particles injected in the fast solar wind cannot reach the forward shock due to the SI acting as a diffusion barrier.
The formation of  an extra population of accelerated particles solely as a consequence of cross-field diffusion illustrates again  how a weak cross-field motion can significantly alter the particle population in the heliosphere when the solar wind is more complex than a simple Parker configuration.  This is particularly important  since in reality, a Parker  configuration is  relatively rare, and can, most of the time, only be found during solar minimum. 

We conclude by noting that the more complex magnetic field configuration of the CIR produced particle time-intensity profiles that differ strongly in shape from the ones obtained when using a nominal IMF. By placing virtual observers at different locations in the heliosphere, we illustrated that the intensity profiles can largely vary from one to the other, even for observers located closely to each other,
that is, separated by $<10^\circ$ in longitude.
These differences can only be attributed to the varying solar wind conditions since the particles  were injected uniformly over the selected region at the inner boundary of EUHFORIA.  
Hence, the background solar wind can have a major influence on the particle transport, which  illustrates the necessity for using more realistic background solar wind configurations when studying SEP events. 
Magnetohydrodynamic codes like EUHFORIA can, to some extent, provide such realistic background winds. 

\begin{acknowledgements}
N. Wijsen is a PhD Fellowship of the Research Foundation Flanders (FWO). The computational resources and services used in this work were provided by the VSC (Flemish Supercomputer Center), funded by the Research Foundation Flanders (FWO) and the Flemish Government – department EWI. The work at University of Barcelona was partly supported by the Spanish Ministry of Economy, Industry and Competitivity under the project AYA2016-77939-P, funded by the European Union's European Regional Development Fund (ERDF), and under the project MDM-2014-0369 of ICCUB (Unidad de Excelencia ‘María de Maeztu’). The work at University of Helsinki was carried out in the Finnish Centre of Excellence in Research of Sustainable Space (Academy of Finland grant numbers 312390 and 312351). The authors thank Neus Agueda, Blai Sanahuja, and Rami Vainio for valuable discussions.
\end{acknowledgements}

\bibliographystyle{aa.bst}

\end{document}